\documentclass[preprint,12pt]{elsarticle}

\usepackage{courier}
\usepackage{amssymb}
\usepackage{amsmath}
\usepackage[htt]{hyphenat}
\usepackage{url}
\usepackage{hyperref}
\usepackage{bm}
\usepackage{booktabs}
\usepackage{multirow}
\usepackage{tikz}
\usepackage{tikz-3dplot}
\usetikzlibrary{calc}
\biboptions{sort&compress}

\newcounter{bla}

\journal{Computer Physics Communications}

\begin{document}

\begin{frontmatter}



\title{RMT: R-matrix with time-dependence. Solving the
 semi-relativistic, time-dependent Schr\"{o}dinger equation for general,
multi-electron atoms and molecules in intense, ultrashort, arbitrarily polarized laser pulses.}


\author[a]{Andrew C. Brown\corref{author}}
\author[a]{Gregory S. J. Armstrong}
\author[b]{Jakub Benda}
\author[a]{Daniel D. A. Clarke}
\author[a]{Jack Wragg}
\author[c]{Kathryn R. Hamilton}
%\author{Michael Lysaght}
\author[d]{Zden\v{e}k Ma\v{s}\'{i}n}
%\author{Robert McGibbon}
%\author{Laura Moore}
%\author{Jonathan Parker}
%\author[d]{Martin Plummer}
%\author{Ken Taylor}
\author[b]{Jimena D. Gorfinkiel}
\author[a]{Hugo W. van der Hart}

\cortext[author] {Corresponding author.\\\textit{E-mail address:} andrew.brown@qub.ac.uk}

\address[a]{Centre for Theoretical Atomic, Molecular and Optical Physics, School of Mathematics and Physics, Queen's University Belfast, Belfast  BT7 1NN, United Kingdom}
\address[b]{School of Physical Sciences, The Open University, Walton Hall, MK7 6AA Milton Keynes, UK}
\address[c]{Department of Physics and Astronomy, Drake University, Des Moines, Iowa 50311, USA}
\address[d]{Institute of Theoretical Physics, Faculty of Mathematics and Physics, Charles University, V Hole\v{s}ovi\v{c}k\'{a}ch 2, 180 00 Prague 8, Czech Republic}

\begin{abstract}
RMT is a program which solves the time-dependent Schr\"{o}dinger equation for general, multielectron atoms, ions and molecules interacting with laser light.  As such it can be used to model ionization (single-photon, multiphoton and strong-field), recollision (high-harmonic generation, strong-field rescattering) and, more generally, absorption or scattering processes with a full account of the multielectron correlation effects in a time-dependent manner. Calculations can be performed for targets interacting with ultrashort, intense laser pulses of long wavelength and arbitrary polarization. Calculations for atoms can optionally include the Breit-Pauli correction terms for the description of relativistic (in particular, spin-orbit) effects.

\end{abstract}


\end{frontmatter}

% \linenumbers

% Computer program descriptions should contain the following
% PROGRAM SUMMARY.

{\bf PROGRAM SUMMARY}
  %Delete as appropriate.

\begin{small}
\noindent
{\em Program Title:} (RMT) R-matrix with time-dependence \\
{\em Licensing provisions:} GPLv3 \\
{\em Programming language:} Fortran                             \\
{\em Program repository available at:} \href{https://gitlab.com/Uk-amor/RMT}{https://gitlab.com/UK-AMOR/RMT}\\
{\em Computers on which the program has been tested:} Cray XC40 BESKOW, Cray XC30 ARCHER, Cray XK7 TITAN, TACC Stampede2, DELL linux cluster, DELL PC \\
{\em Number of processors used:} Min. 2, Max. tested 16,416\\
{\em Number of lines in program:} 25,247\\
{\em Distribution format:} git repository\\

\noindent{\em Nature of problem:}\\
  The interaction of laser light with matter can be modelled with the time-dependent Schr\"odinger equation (TDSE). The solution of the TDSE for general, multielectron atomic and molecular systems is computationally demanding, and has previously been limited to either particular laser wavelengths and intensities, or to simple, few-electron cases. RMT overcomes this limitation by using a general approach to modelling dynamics in atoms and molecules which is applicable to multi-electron systems and a wide range of perturbative and non-perturbative phenomena.\\
  
\noindent{\em Solution method:}\\
  We use the R-matrix paradigm, partitioning the interaction region into an `inner' and an `outer' region. In the inner region (within some small radius of the nucleus/nuclei), full account is taken of all multielectron interactions including electron exchange and correlation. In the outer region, far from the nucleus/nuclei, these are neglected and a single, ionized electron moves in the long-range potential of the residual ionic system and the laser field. The key computational aspect of the RMT approach is the use of a different numerical scheme in each region, facilitating efficient parallelization without sacrificing accuracy. Given an initial wavefunction and the electric field of the driving laser pulse, the wavefunction for all subsequent times and the associated observables are computed using an explicit, Arnoldi propagator method. \\
  
\noindent{\em Additional comments including restrictions and unusual features:}\\
The description of the atomic/molecular structure is provided from other,
time-independent R-matrix codes \cite{connorb, rmatrix_ii_repo, UKRMOL+}, and the capabilities (in terms of structure) are, in some sense, inherited therefrom.  Thus, the atomic calculations can optionally include Breit-Pauli relativistic corrections to the Hamiltonian, in order to account for the spin-orbit effect.  However, no such capability exists for the molecular case. Furthermore,  the fixed-nuclei approximation is adopted in the molecular calculations (so nuclear motion is neglected). Similarly, all calculations are restricted to the description of a single electron in the outer region, and consequently the study of  double-ionization phenomena is not yet within the capabilities of the method. Finally, the parallel strategy employed necessitates the use of at least two (and usually many more) computer cores. As a result, there is no option for serial calculations and, for most realistic cases, a massively parallel architecture (several hundred cores) will be required.  \\

% items referenced in program summary only:

\end{small}

\section{Introduction}
\label{sec:introduction}
% Some information on strong field physics
The last three decades have seen a rapid development in laser technologies. As well as large-scale synchrotron and free-electron laser facilities, offering femtosecond pulses with high photon energies and fluxes, the development of table-top, high-harmonic generation (HHG) sources has enabled attosecond light pulses for ultrafast experiments. The science enabled by these advances has shifted focus away from electronic {\it structure} and towards electronic {\it dynamics}, as the time-dependent behaviour of the systems under inspection becomes resolvable with sufficiently-short, sufficiently-precise laser pulses.

Concurrently with these technological developments, a shift in theoretical approaches has taken place. The seminal work of Paul Corkum \cite{corkum1993} in the early 1990s introduced the so-called `simple-man's' or `three-step' semi-classical model of electron dynamics in a strong-laser field, and inspired the next two decades of theoretical atomic, molecular and optical physics. Indeed, if the complicated, correlated multielectron dynamics of the quantum picture could be replaced with a massively simplified model, which neatly captured all of the important experimental features, why should one bother with the former? It is only in the last several years that experimental techniques have become sufficiently advanced to probe the dynamics of correlated electrons, necessitating a corresponding shift in our theoretical capabilities.

% Other methods / state of the art
Several computational methods have been proposed and implemented in this vein - methods which are capable of including some description of multielectron interactions within the framework of a time-dependent, strong-field approach. Early techniques used the Floquet ansatz to treat the electric field as effectively infinite in duration, precluding the need for a truly time-dependent method \cite{RMF1,RMF2}. This, however, became unrealistic as laser-pulse durations became ever shorter. The {\sc{helium}} code was one of the first to solve the time-dependent Schr\"{o}dinger equation for a two-electron system with a full account of dielectronic interactions, but the method was not easily extended to general, multielectron systems \cite{HELIUM}. In the early 2000s, the multi-configuration, time-dependent Hartree Fock (MCTDHF) approach \cite{MCTDHF}, permitted a good description of multielectron effects, but was not, until recently, applicable to laser regimes beyond the few-photon limit \cite{TDRASSCF-multielectron_atoms, TDCASSCF}. The time-dependent R-matrix (TDRM) method was developed with the explicit aim of describing general, multielectron atoms and ions in ultrashort, intense laser pulses. This goal was achieved, and a number of successful applications of the method demonstrated the importance of multielectron effects in strong-field processes \cite{collect_c+2,collect_c+_m1,brown_prl, brown_m1, brown_helium}. 

The TDRM method was encumbered, however, with the need to propagate the R-matrix throughout an expansive region of configuration space, a computationally expensive task which limited the application of the code to scenarios in which short-wavelength (extreme-ultraviolet or visible) laser light was employed. For this reason, methods such as the time-dependent configuration-interaction singles (TDCIS) approach \cite{TDCIS} led the way in theoretical support for experiments which typically employed IR (800~nm) wavelengths. TDCIS theory was employed successfully to describe the giant resonance in the high-harmonic spectrum of Xe \cite{TDCIS_Xe_giant_resonance}, the effect of spin-orbit coupling on HHG \cite{TDCIS_spin-orbit}, as well as the attosecond transient-absorption spectroscopy (ATAS) of Kr \cite{synthesised_light_transients}. The technique is, however, restricted to the description of single electronic excitations, and to closed-shell systems (in particular, noble-gas atoms). As such, a more flexible method which can fill these niches is required.

% the need for our method
The computational difficulty in describing electronic dynamics in a strong laser field is two-fold. First, the combination of a low photon energy with high intensity stimulates the absorption of multiple photons.  As each absorption can increase the angular momentum of the system, a correct description of the dynamics requires a prohibitive number of angular-momentum terms to be included, creating substantial computational overhead. Secondly, ionized electrons can be steered by a strong field in a classical trajectory which re-encounters the residual ion (this is the centrepiece of Corkum's three-step model \cite{corkum1993}). The maximum radial displacement of a returning electron in a strong field is proportional to the squares of the wavelength and intensity, and so an accurate description requires prohibitively large interaction regions. 

The R-matrix with time-dependence method, first introduced in 2008 for Hydrogen
\cite{RMT_hydrogen}, and then in 2011 for the non-relativistic dynamics of
general, multielectron atoms and ions in linearly polarized laser
pulses \cite{RMT}, addressed these problems by combining aspects of the TDRM and {\sc
helium} approaches. In particular, the use of an R-matrix basis affords an accurate treatment of short-range, multielectron correlations in proximity to the nucleus, while the application of {\sc helium} finite-difference techniques permits an effective description of the radial motion of the ejected electron. Importantly, these distinct numerical approaches benefit from rather natural modes of parallelization, which in RMT are combined in an efficient and scalable implementation to address the key computational difficulties of modelling strong-field dynamics. The RMT method has been applied to many cutting-edge problems in strong-field physics, including the experimentally relevant techniques of HHG \cite{ola_nearIR, brown_xuvhhg, clarke_xihhg, hamilton_two_colour}, ATAS \cite{RMT_ATAS} and strong-field rescattering \cite{ola_rescattering}. 

On the molecular side several  {\it ab initio} approaches have been implemented to describe interaction with strong fields, with the molecular R-matrix Floquet approach~\cite{burke_floquet_2000,RMF_H2}  one of the first methods applicable to multi-electron systems. The implementation was limited to diatomic molecules and monochromatic (i.e. long) fields.% but the one-electron wavefunctions used the Slater functions to describe the bound orbitals and numerical functions were used for the continuum description.

Subsequently several different approaches were developed including the
\texttt{haCC}~method of Majety and Scrinzi~\cite{majety_2015} that has been
applied to diatomic and triatomic molecules~\cite{majety_prl,majety_co2_2017},
the Spanner-Patchkovskii approach~\cite{spanner_2009} with approximate exchange
applicable to small and medium-sized molecules~\cite{spanner_2013} and the
B-Spline Algebraic Diagrammatic Construction (ADC) of Averbukh and
Ruberti~\cite{ruberti_2014,ruberti_2018, ADC}. Recently, the XChem quantum chemistry package has been extended~\cite{Marante_code_madrid} to support calculations of field-ionization of small molecules~\cite{Madrid_N2} (and atoms). For the smallest molecules like H$_2$ a near full quantum (including nuclear motion) treatment in perturbative fields has been developed~\cite{palacios_2015}. In larger molecules (e.g. CH$_4$) interacting with strong fields, the coupled electronic-nuclear problem can be solved using the multiconfigurational strong-field approximation with Gaussian nuclear wave packets method~\cite{patchkovskii_2017}. %Finally, the continuum functions obtained from DFT codes~\cite{Decleva_DFT} have been used in a variety of strong-field applications ranging from small to large molecules~\cite{nisoli_2017}.
Extending the use of the RMT method to molecules was a natural step to enable calculations of a similar high quality to those that can be performed for atoms. We note that an alternative implementation of RMT for one-electron systems has recently been demonstrated for both atomic Hydrogen and the molecular Hydrogen ion \cite{nikolopoulos_mol_rmt}.

The key extensions of the original atomic RMT approach, which are now implemented in the RMT codes and discussed in this paper, afford the description of 

\begin{itemize}
    \item 
    the dynamics of atoms and atomic ions in arbitrarily polarized laser pulses,
    \item
    the dynamics of molecules and molecular ions in such pulses, as well as
    \item
    spin-orbit corrections to the atomic structure.
\end{itemize}

In this paper, we describe the computational implementation of these extensions. Additionally, we highlight the features of the RMT codes which enable more substantial calculations, necessitating enhanced parallelism and memory management.

\section{Overview of R-matrix approach}
% Simple sketch of basic R-matrix principles, including:
A detailed presentation of the RMT method can be found in Ref. \cite{RMT}. Here, we provide only a brief overview. RMT theory offers an {\it ab initio} and non-perturbative technique for solving the time-dependent Schr\"{o}dinger equation (TDSE), appropriate to general atomic and molecular systems exposed to intense and ultrashort laser pulses. The method employs the traditional R-matrix paradigm of dividing configuration space into several distinct regions. In particular, RMT treats two such regions. 

%A detailed presentation of the RMT method can be found in Ref. \cite{RMT}. Here, we provide only a brief overview. The RMT approach provides a numerical solution of the time-dependent Schr\"{o}dinger equation (TDSE) for general atomic and, most recently, molecular systems exposed to intense and ultrashort laser pulses. The fundamental concept of the R-matrix paradigm is the division of space into several distinct regions: RMT employs a two-region approach. 

% division of space
Consider an ($N+1$)-electron system (atom, ion or molecule) interacting with a laser pulse. In an `inner region', confined to small radial distances from the target nucleus, the system is treated with a full account of all electronic interactions, including electron exchange and correlation by means of a configuration-interaction approach. Should an electron escape this inner region, it becomes spatially isolated from the residual $N$ electrons, and electron exchange may be neglected. In this `outer region', a single, ionized electron moves only under the influence of the laser field and the long-range, multipole potential of the $N$-electron residue. Note that whilst the inner region carries the full complexity of the many-body problem, the outer region entails an effectively one-body analysis. The computational simplicity afforded by such a treatment is especially valuable for modelling strong-field processes, where the ionized electron may travel far from the nucleus, and thus the wavefunction must be propagated throughout expansive spatial regions and for long periods of time.

%This simplification of the description in the outer region allows for a tractable solution of the TDSE to be determined, as in principle the electron may travel far from the nucleus, and the outer region can be (otherwise) prohibitively large.

% R-matrix basis
Key to the accurate determination of multielectron effects is a judicious choice of basis for the wavefunction. Both the atomic and molecular bases are constructed by treating the ($N+1$)-electron system as an $N$-electron residual ion, plus an additional electron. This additional electron can occupy a bound orbital - to construct the ground or excited states - or it may be in the continuum, offering a good description of the ionized, $N$-electron residue and its interaction with the outgoing electron. 

% channel functions
This choice of basis also lends itself to a clear construction for the outgoing-electron wavepacket, which can be thought of as travelling in electron emission `channels'. The channel description captures the essential asymptotic properties of the outgoing electron and the residual ion, so that the channel wavefunctions and populations can be used to predict experimental observables such as photoelectron energy and momentum spectra, ionization rates and photoionization cross sections.
This channel formalism also affords a description of
spin-multiplicity as residual-ion states of differing multiplicities
may be selectively included in the calculations. Thus, even though the total
spin of the system is conserved, the outgoing electron may couple to various
spin states of the residual ion. (We note that the spin multiplicity does not
appear explicitly in the numbering convention used for the channels (see
Sec. \ref{channel-populations}) but is manifest in the energy of the residual
ion state.)

The key to the success of RMT over previous implementations of time-dependent R-matrix theory \cite{tdrm} is its use of a different numerical scheme in each region. In the inner region, accurate and efficient determination of the multielectron wavefunction is ensured by the use of a $B$-spline basis (mixed with Gaussians in the case of molecular calculations), whereas a grid-based, finite-difference approach is employed in the outer region, which also facilitates enhanced parallelism. At variance with more traditional R-matrix approaches, the wavefunction is matched explicitly at the interface of the two regions, rather than via an R-matrix. 

A multilayered parallel implementation (see Sec. \ref{parallelization-strategy}) is key to the previous and ongoing success of the RMT approach. Continued expansion of its capabilities, and application to ever more complex strong-field problems, is only possible through an efficient mapping of large computational loads to correspondingly large core counts.

\subsection{Arbitrarily polarized light fields}
\label{sec:arbpol}
% just need a sketch here- can reference PRA article to provide bulk of theory. 
The treatment of atomic dynamics in arbitrarily polarized laser light is a substantial computational task. It requires lifting symmetry restrictions allowed for linearly polarized fields, which limited the scale of previous RMT calculations. For a field linearly polarized along the $z$ axis, the total orbital magnetic quantum number $M_L$ is conserved, and the TDSE may be solved using a set of dipole-accessible $LS\pi$ symmetries. However, with a laser field of arbitrary polarization, conservation of $M_L$ is lost: even in the dipole approximation, such fields induce transitions in which $M_L$ must change by $\pm1$. 

Therefore, in the atomic case, the main computational difficulty involves the incorporation of magnetic sublevels, so that the TDSE may be solved using $LM_LS\pi$ symmetries. For a given maximum angular momentum $L_{\rm max}$, the number of such symmetries $N_{\rm sym}$  is
\begin{equation}
    N_{\rm sym} = 2(L_{\rm max}+1)^2 .
    \label{maxnsym}
\end{equation}
%

% DDAC drafting.....
% A detailed discussion of the theoretical and computational developments required for arbitrary light fields is given in Ref. \cite{RMT_arb}, and so we merely provide a brief summary here. The main computational difficulty arises the replacement of each $LS\pi$ symmetry with $(2L + 1)$, $LM_LS\pi$ symmetries, which carry their own atomic-structure and dipole-coupling data. For a given maximum angular momentum $L_{\rm max}$, whose choice is largely dictated by the laser intensity and wavelength, the number of such symmetries $N_{\rm sym}$ is....Note the quadratic scaling with $L_{\rm max}$, which contrasts with the linear one found in the case of pure, linear polarization along the $z$-direction.

This increase in scale is felt in both regions: the inner region must handle a large number of symmetries (and, correspondingly, dipole blocks) while the outer region entails a large number of coupled channels. This has significant implications for code parallelization, detailed in Sec.~\ref{rmt-code-computational-considerations}. 

\begin{figure}
    \centering
    \centerline{\includegraphics[width=0.8\textwidth]{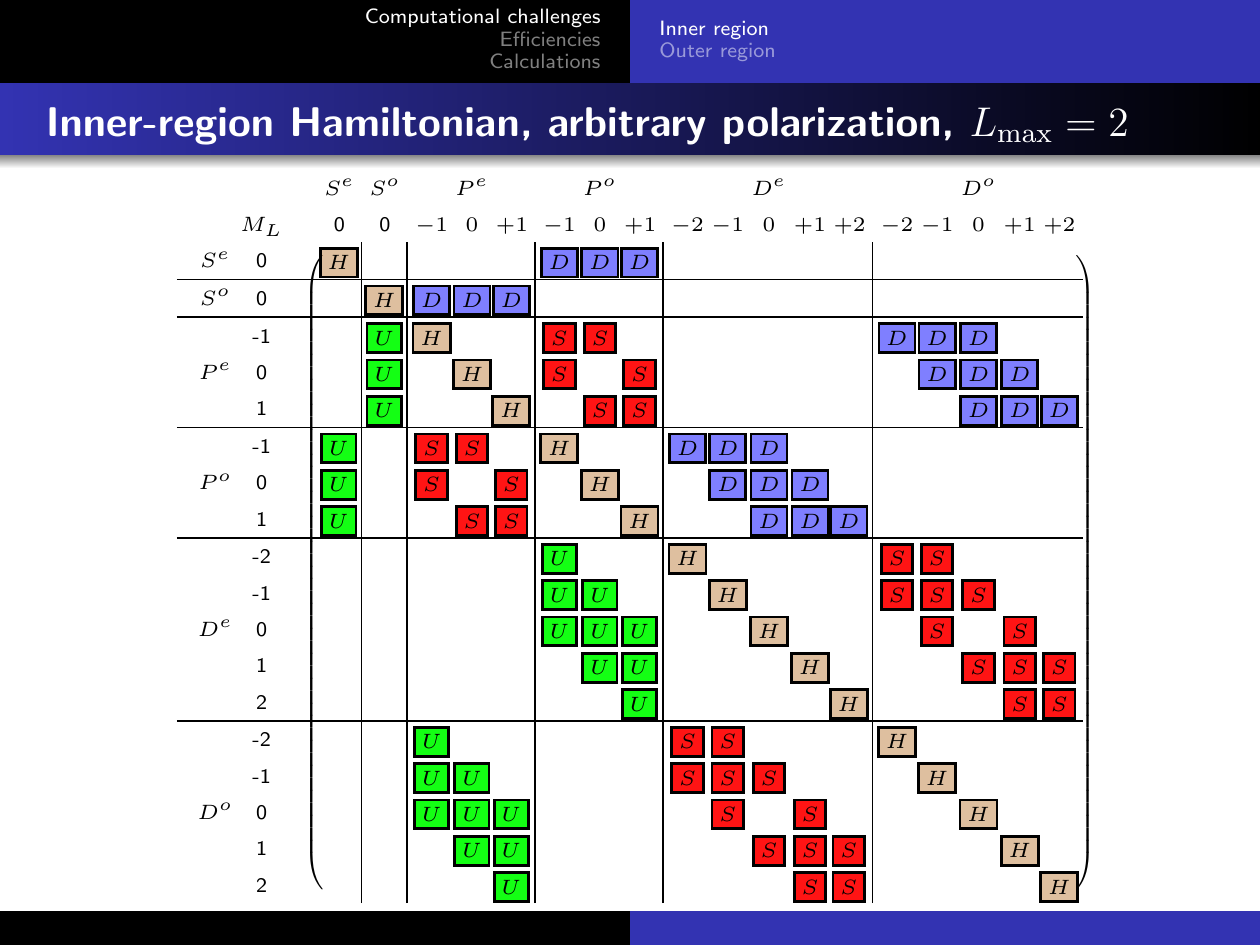}}
    \caption{Inner-region Hamiltonian for $S$, $P$ and $D$ symmetries, assuming an $S^e$ initial state. Dipole blocks labelled $D$, $S$ and $U$ (`down', `same' and `up') indicate $\Delta L=-1,0,1$ transitions.}
    \label{innerham}
\end{figure}

The scale of the development is clearly manifest in the structure of the inner-region Hamiltonian, shown in Fig.~\ref{innerham} for all $LM_LS\pi$ symmetries up to $L_{\rm max}=2$. Each symmetry can couple to up to 9 others under the selection rules $\Delta L=0,\pm1$, $\Delta M_L = 0,\pm1$. 

The structure of the Hamiltonian shown in Fig.\;\ref{innerham} depends on the laser polarization. In RMT, the default polarization plane is the $zy$ plane, such that a general, elliptically polarized electric field takes the form
\begin{equation}\label{def:efield}
    {\bf E}(t) = F(t){\rm Re}\left[\hat{\bf e}
    \;e^{-i(\omega t + \varphi)}\right],
\end{equation}
where  $\hat{\bf e}=(\hat{\bm\epsilon}+i\eta\hat{\bm\zeta})/\sqrt{1+\eta^2}$ is an arbitrary polarization vector, $\eta$ is the ellipticity, and $\varphi$ is a carrier-envelope phase. By default, $\hat{\bm\epsilon} = \hat{\bf z}$ is the major axis of the polarization ellipse, and $\hat{\bm\zeta} = -\hat{\bf y}$ is the minor axis. This choice allows calculations for linear fields ($\eta = 0$) to be polarized along the $z$ axis as usual. To use alternative polarization planes, a set of Euler angles may be specified on input, which permit a solid rotation of the polarization plane away from from the $zy$ plane (See Sec.~\ref{sec:input}).

\begin{figure}
    \centering
    \centerline{\includegraphics[width=0.8\textwidth]{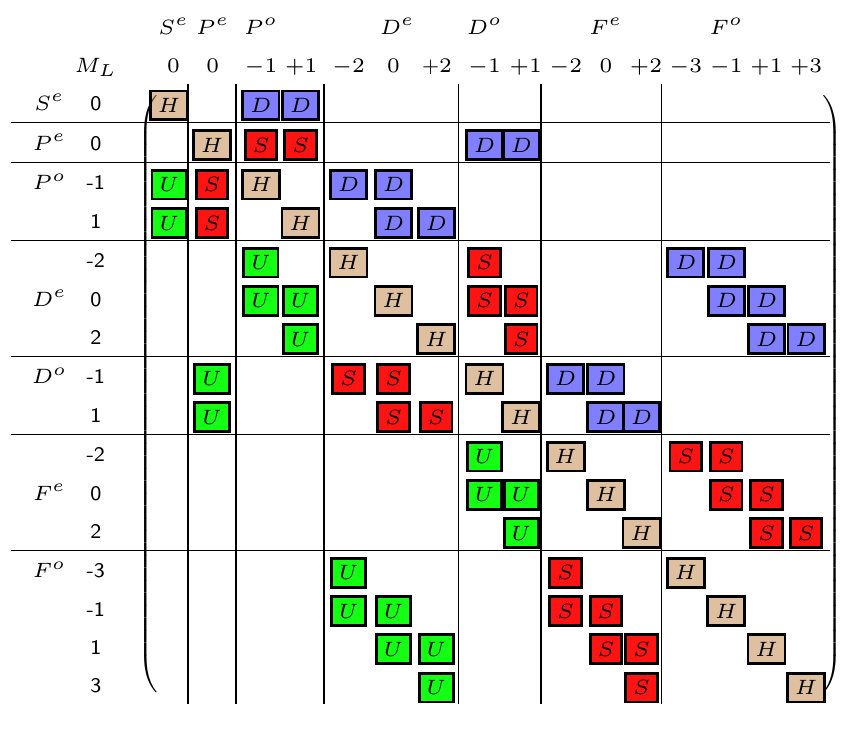}}
    \caption{Inner-region Hamiltonian for $S$, $P$, $D$ and $F$ symmetries, accessible for $xy$-plane polarization, assuming an $S^e$ initial state.  Dipole blocks labelled $D$, $S$ and $U$ indicate $\Delta L=-1,,0,+1$ transitions.}
    \label{xyham}
\end{figure}

One special case is worth highlighting, namely the choice of $xy$-plane laser polarization. This computationally efficient case is enabled in RMT calculations by setting {\texttt{xy\_plane\_desired = .true.}} on input. For this choice of polarization, $\Delta M_L=0$ transitions are forbidden, and so only a subset of all magnetic sublevels are dipole-accessible. In fact, only half of all sublevels are accessible, so that the number of symmetries required is given by
\begin{equation}
    N_{\rm sym} = (L_{\rm max}+1)^2 .
\end{equation}
The reduction (relative to \(N_{\rm sym}\) in Eq.\;\eqref{maxnsym}) is made clear in the structure of the Hamiltonian for $xy$-plane polarization, as shown in Fig.\;\ref{xyham} for \(L_{\rm max}=3\). Here, only half of all \(M_L\) values must be included, and the elimination of $\Delta M_L=0$ transitions means that each symmetry can couple to a maximum of 6 others. 

It is also possible to limit the number of symmetries by only retaining $M_L$ values within a restricted range, such that $|M_L|\leq M_{L_{\rm max}}$. For a given value of $M_{L_{\rm max}}$, the number of symmetries is given by
\begin{equation}
    N_{\rm sym} = 2\left[
                    (M_{L_{\rm max}}+1)^2 
                    + 
                    (2M_{L_{\rm max}}+1)(M_{L_{\rm max}}-L_{\rm max}) 
                    \right].
\end{equation}
This formula applies only if the electric field has a non-zero $z$-component, since it accounts for symmetries that can be coupled by $\Delta M_L=0,\pm1$ transitions. In particular, it may be used for problems in which the \(z\) component of the laser field dominates, driving \(\Delta M_L=0\) transitions strongly, with weaker \(x\) and/or \(y\) components driving \(\Delta M_L=\pm1\) transitions.

Since the major computational task in the inner region is the handling of a large number of symmetry blocks, this aspect of the calculation is parallelized, as demonstrated in Sec.\;\ref{inner-region-parallelization}. Therefore, the options outlined above have significant impact on the computational resources necessary for the inner region.

In the outer region, the multielectron wavefunction is given as a standard close-coupling expansion, involving the time-dependent reduced radial wavefunction of the ejected electron in each channel, and channel functions that handle all other degrees of freedom. The resulting TDSE for the reduced radial wavefunctions is then solved using a finite-difference discretization.

A detailed discussion of the theoretical development required for this extension is given in Ref.\;\cite{rmt_arb} for the atomic case and described in~\ref{sec:appA} for the molecular case.   

\subsection{Relativistic corrections}

% extensions relevant to relativistic input
Recent work has enabled the application of RMT to the laser-driven dynamics of
heavier atomic systems, in which the spin-orbit interaction cannot be neglected.
This has been achieved by extending the RMT code to read atomic-structure data
generated within a $jK$ coupling scheme, such as that supplied by the RMatrixI
suite \cite{belfastatomic}.

Using the $jK$ coupling scheme allows for states with half-integer quantum numbers, but for calculations performed in standard $LS$ coupling, the quantum numbers are stored in integer form (note that spin is identified by spin multiplicity). When considering $jK$ coupling data, every quantum number that could be half-integer is doubled (thus, $J$ is stored as $2J$, $M_{J}$ is stored as $2M_{J}$, etc.), which maintains unique identification of the quantum numbers while mitigating the need for non-integer storage. 

The main adjustments necessary to describe the spin-orbit effects are in the input data provided to RMT: the Hamiltonian and dipole matrix elements are constructed by the RMatrixI suite using $jK$ coupling incorporating the spin-orbit interaction.  Once the input data is read in and stored appropriately, RMT's scheme for describing the system in terms of symmetries (inner region) or channels (outer region), coupled in accordance with dipole selection rules, can be applied just as well to the non-relativistic or semi-relativistic cases.

%The spin-orbit splitting itself is handled within the input data, specifically in the inner region and target-state eigenvalues, and the long-range potential matrices. Otherwise, no changes to RMT are necessary to represent the spin-orbit splitting. Put differently, once the input data is read in and stored appropriately, RMT's scheme for describing the system in terms of symmetries (inner region) or channels (outer region), coupled in accordance with dipole selection rules, can be applied just as well to the non-relativistic or semi-relativistic cases.

In principle it is possible to run RMT for atomic data described in $jK$ coupling, but which does not include the spin-orbit interaction. In this case, and for a fixed choice of target, the resulting wavefunction should be identical to that created from an $LS$-coupling data set (after the channel recoupling has been taken into account). This may be used to test the appropriateness of the $LS$ coupling scheme for any given atomic structure model (see Sec.~\ref{sec:results}). 

\subsection{Molecular calculations}

RMT has been newly extended to support molecular targets: the method itself is virtually unchanged with respect to the atomic case. The molecular structure data and all other quantities required for the time-propagation are provided by the UKRmol+ package \cite{UKRMOL+} that implements the time-independent R-matrix method for molecules \cite{UKRMOL1}. The most apparent difference is that while the (non-relativistic) electronic wavefunction of an atom can be expanded in total angular momenta \(L\) and their projections \(M_L\) due to the atom's spherical symmetry, this is not possible in the molecular case. Instead, the total molecular wavefunction is written as a sum of up to 8 electronic wavefunctions, each transforming according to an irreducible representation of the D$_{2h}$  point group or one of its subgroups  (e.g., C$_2$, C$_{2v}$, etc.) depending on the molecular symmetry. In the case of linear molecules, the projection $M$ of the angular momentum along the molecular axis is a conserved quantity, but its incorporation requires the use of non-Abelian point groups which, similarly to most quantum chemistry software packages, are not implemented in UKRmol+. The selection rules and the block structure of the Hamiltonian are given by the point group to which the molecule, and the corresponding irreducible representations to which the contributing wavefunctions, dipole operator components, etc., belong.

\begin{figure}
    \centering
    \centerline{\includegraphics[width=0.8\textwidth]{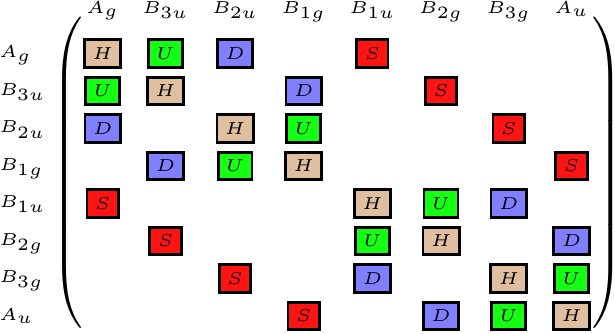}}
    \caption{Inner-region Hamiltonian for the case of a molecule belonging to the $D_{2h}$ point group and electric field having all three ($x$, $y$, $z$) components non-zero. Dipole blocks labelled $D$, $S$ and $U$ indicate dipole transitions induced by the $x$, $y$, $z$ field components respectively.}
    \label{innerham-mol}
\end{figure}

Therefore, in the molecular case, the structure of the Hamiltonian matrix is limited to a maximum of 8-by-8 blocks for the case of the highest-symmetry D$_{2h}$ point group (see Fig.~\ref{innerham-mol}). All molecular complexity then manifests only in the size of these blocks. In molecular calculations, the index of the irreducible representation is used in place of \(L\) to label the Hamiltonian blocks; no projection \(M_L\) is defined. In the dipole approximation, the electric field then couples a given irreducible representation to at most three others, as given by the appropriate group multiplication table. Additional technical differences between the atomic and the molecular RMT calculations are described in~\ref{conventions}.

%It relies on providing the
%\begin{itemize}
% \item energies of both the \(N\)-electron and \((N + 1)\)-electron states,
% \item dipole matrix elements between the inner-region \((N + 1)\)-electron states,
% \item dipole matrix elements between the \(N\)-electron (ionized) states,
% \item mapping between the inner-region \((N + 1)\)-electron states and the outer-region target~+~projectile channels around the R-matrix sphere boundary
%\end{itemize}
%and some further data needed for evaluation of the outer region potentials. 

%The second notable dissimilarity is a different channel expansion of the wavefunction in the outer region,
%\begin{equation}
%    \Psi^{\Gamma}(\bm{r}_1,\dots,\bm{r}_{N+1}) = \frac{1}{r_{N+1}} \sum_i \Phi_i^{\Gamma_i}(\bm{r}_1,\dots,\bm{r}_N) F_i(r_{N+1}) Y_{l_i m_i}(\hat{\bm{r}}_{N+1}) \,,
%\end{equation}
%which makes direct use of the relevant group multiplication to compose the wavefunction of the required total symmetry \(\Gamma\); no additional summation weighted by coupling coefficients is needed here.

\section{Input data}
\label{sec:input}
\subsection{Atomic/Molecular Data}
The input data containing information about the ground and excited states of the target system, the dipole couplings and the atomic $B$-spline basis (or the amplitudes of the $(N+1)$-electron wavefunctions in the molecular case) are produced by one of three different software packages. Non-relativistic atomic data can be provided by either the R-matrix I \cite{connorb} or R-matrix II packages \cite{RMT_repo}. Semi-relativistic data is produced by R-matrix I, and molecular data by UKRmol+ \cite{UKRMol+_repo}, all of which can be obtained, along with documentation, at the repositories linked.

The input files required are slightly different depending on the package used.  For molecular calculations, all relevant input data is stored in a single file {\texttt{molecular\_data}}. For both atomic cases, the Hamiltonian data is stored in file {\texttt{H}}, and the spline basis information in files {\texttt{Splinedata}} and {\texttt{Splinewaves}}. The dipole information from R-matrix II is stored in files {\texttt{d}} and {\texttt{d00}}. For R-matrix I input, an individual file is used for each dipole-coupling block. Hence the header information is stored in file {\texttt{D00}} and the individual dipole files are of the form {\texttt{D\#\#\#}} (e.g., {\texttt{D001}}, {\texttt{D002}}, etc.). We note that the somewhat confusing naming of these files is set by other codes, and the decision not to rename them ensures continued compatibility therewith.  

\subsection{Runtime calculation parameters}
All calculation parameters are set at runtime using the namelist {\texttt{\&InputData}} written in the file {\texttt{input.conf}}, which should be located in the root directory. Several sample input files can be found in the {\texttt{./tests/}} directory of the repository. The general format is

\vspace{12pt}
\noindent \texttt{\&InputData \\ <parameter\_1\_name> = <parameter\_1\_value> \\ <parameter\_2\_name> = <parameter\_2\_value> \\ $\ldots$ \\  / }
\vspace{12pt}

\subsection{Required parameters}
In order to execute, each calculation requires at least the following parameters to be defined in the file {\texttt{input.conf}}.

\begin{itemize}
    \item 
    {\texttt{no\_of\_pes\_to\_use\_inner }(INTEGER)} $N_{\mathrm{inn}}$ defines the number of cores to be used for the inner-region calculation. This must be greater than or equal to the number of inner-region Hamiltonian blocks. See Sec. \ref{rmt-code-computational-considerations}.
    \item
    {\texttt{no\_of\_pes\_to\_use\_outer} (INTEGER)} $N_{\mathrm{out}}$ defines the number of cores to be used for the outer-region calculation. 
    \item 
    {\texttt{x\_last\_others}} (INTEGER) $N_x$ defines the number of grid points per outer-region core. Due to the finite-difference method employed, this value must be a multiple of 4, and greater than or equal to twice the value of \texttt{taylors\_order}, the order of the propagator used. 
    \item
    {\texttt{x\_last\_master}} (INTEGER) $N_m$ defines the number of grid points for the first outer-region core. Due to the finite-difference method employed, this value must be a multiple of 4, and greater than or equal to twice the value of \texttt{taylors\_order}. To achieve good load-balancing, this value should usually be chosen to be less than {\texttt{x\_last\_others}}, as the outer-region master has additional work to perform.
    \item
    {\texttt{deltaR}} (REAL) $\delta r$ defines the outer-region grid spacing in atomic units. In practice,  this means that the outer region will be of size
    \begin{equation}
     r_{\mathrm{out}} = (N_{\mathrm{out}}-1)\; N_x \; \delta r + N_m \; \delta r.
    \end{equation}
    \item
    {\texttt{steps\_per\_run\_approx}} (INTEGER) $N_s$ defines the number of time steps to be executed in the propagation. 
    \item
    {\texttt{final\_T}} (REAL) $t_f$ defines the time, in atomic units, up to which the wavefunction solution should be propagated. Implicitly, then, the time step for the propagation is defined by
    \begin{equation}
        \delta t = \frac{t_f}{N_s},
        \label{dt_define}
    \end{equation}
    and it should be ensured that this quantity is sufficiently small to give convergence of the numerical results. In practice, we have found that a time step of $0.01$~a.u. gives good results for most applications. 
\end{itemize}

\subsection{Optional Parameters}
Additional parameters which can be defined for any given calculation are outlined below. The type and default value of each parameter is given in brackets.

\subsubsection{Target information}
\begin{itemize}
    \item 
    {\texttt{molecular\_target}} (LOGICAL, .false.) defines if the calculation is for an atomic or molecular target. By default, the code assumes atomic.
    \item
    {\texttt{GS\_finast}} (INTEGER, 1) defines which of the included symmetries pertains to the ground state. In the case of molecules, it indicates the irreducible representation to which this state belongs (the order of the irreducible representations for Abelian point groups is described in Ref. \cite{UKRMOL+}). 
    \item
    {\texttt{adjust\_GS\_energy}} (LOGICAL, .false.) defines whether or not the ground-state energy should be adjusted.
    \item
    {\texttt{GS\_energy\_desired}} (REAL, 0.0) defines the desired ground-state energy.
    \item 
    {\texttt{dipole\_format\_id}} (INTEGER, 2) defines which version of the R-matrix codes has been used to generate the input data. The default is the R-matrix II codes (2). For relativistic calculations, R-matrix I (1) will be used. For molecular calculations, this parameter is ignored.
     \item 
    {\texttt{coupling\_id}} (INTEGER, 1) defines which coupling scheme to use. The default is $LS$ coupling (1); $jK$ coupling (2) can also be employed. For molecular calculations, this parameter is ignored.
    \item
    {\texttt{lplusp}} (INTEGER, 0) defines the sum, modulo 2, of the angular momentum and parity of the ground state. For instance, in the $^1S^{e}$ ground state of Ne, we have $L=0$ and $\pi = 0$, so that {\texttt{lplusp}}$=0$. For the $^3P^e$ ground state of C, we have $L=1$ and $\pi=0$, so that {\texttt{lplusp}}$=1$. For molecular calculations, this parameter is ignored.

    \item
    {\texttt{ML\_max}} (INTEGER, -1) The maximum absolute value of $M_L$ to be used in the calculation. For molecular calculations this parameter is ignored. 
    For calculations where the $z$-component of the laser electric field is dominant, the emission of electrons with $m_l \neq 0$ can be substantially less probable than those with $m_l=0$. Thus, while we would expect population in states with high angular momenta $L$, we might expect substantially less population in states with high $M_L$. If {\texttt{ML\_max}} is not set in the input file, calculations will retain all magnetic substates, and thus by default $M_{L_{\mathrm{max}}} = L_{\mathrm{max}}$. For large $L_\mathrm{max}$, this can lead to prohibitively large calculations, especially given the requirement of at least one processor core per symmetry block. It is therefore possible to set a limit by providing a value for $M_{L_{\mathrm{max}}}$ in the input file. Naturally, calculations with the laser pulse linearly polarized in the $z$-direction should employ $M_{L_{\mathrm{max}}} = 0$ for maximum efficiency.
\end{itemize}

\subsubsection{Laser parameters}
\begin{itemize}
    \item
    {\texttt{use\_2colour\_field}} (LOGICAL, .false.) allows the use of two, independently controllable laser pulses. 
\end{itemize}
For historical reasons, the primary laser pulse is referred to here and in the code as the IR, the secondary pulse as the XUV. The parameters below thus refer to the primary (IR) pulse.
\begin{itemize}
    \item
    {\texttt{frequency}} (REAL, 0.0) the carrier frequency of the primary laser pulse in atomic units.
    \item
    {\texttt{periods\_of\_ramp\_on}} (REAL, 0.0)
    \item
    {\texttt{periods\_of\_pulse}} (REAL, 0.0)
    The pulse profile has a sine-squared envelope. {\texttt{periods\_of\_ramp\_on}} sets the number of cycles to be used in both the ramp-on and -off parts of the pulse. {\texttt{periods\_of\_pulse}} then sets the total number of cycles (including the ramp-on and -off). Thus 
    \begin{equation} 
    \mbox{\texttt{periods\_of\_pulse}} \geq 2 \times \mbox{\texttt{periods\_of\_ramp\_on}}
    \end{equation}
    Note that the code supports non-integer numbers of cycles.
    \item
    {\texttt{ellipticity}} (REAL, 0.0) The ellipticity of the primary pulse. 
    The default orientation of the polarization plane is such that the major axis is $z$, the minor axis is $y$, and propagation is along the $x$ direction. Therefore, an ellipticity of +1 specifies circular polarization in the $zy$ plane, with positive sense with respect to the +$x$ direction, while an ellipticity of $-1$ corresponds to circular polarization in the same plane, but with positive sense with respect to the $-x$ direction. A value of 0 yields linear polarization, and some value between $-1$ and +1 sets an arbitrary level of ellipticity. By default, the primary pulse is linearly polarized along the $z$ axis. Note that alternative polarization planes may be accessed using the Euler-angle inputs \mbox{\texttt{euler\_alpha}}, \mbox{\texttt{euler\_beta}}, and \mbox{\texttt{euler\_gamma}} (see below).
    \item
    {\texttt{ceo\_phase\_deg}} (REAL, 0.0) The carrier envelope phase measured in degrees.
    \item
    {\texttt{intensity}} (REAL, 0.0) The peak intensity of the primary pulse in units of $10^{14}\mbox{ Wcm}^{-2}$.
\end{itemize}
The equivalent parameters for the secondary (XUV) pulse are: 
\begin{itemize}
    \item
    {\texttt{frequency\_XUV}} (REAL, 0.0) 
    \item
    {\texttt{periods\_of\_ramp\_on\_XUV}} (REAL, 0.0)
    \item
    {\texttt{periods\_of\_pulse\_XUV}} (REAL, 0.0)
    \item
    {\texttt{ellipticity\_XUV}} (REAL, 0.0) 
    \item
    {\texttt{ceo\_phase\_deg\_XUV}} (REAL, 0.0) 
    \item
    {\texttt{intensity\_XUV}} (REAL, 0.0) 
\end{itemize}

\noindent The temporal relationship between the two laser pulses is determined by:
\begin{itemize}
    \item 
    {\texttt{time\_between\_peaks\_in\_fs}} (REAL, 0.0) The time between the central peaks of the primary and secondary pulses measured in femtoseconds. Negative delay corresponds to the secondary (XUV) pulse peak arriving first.
    \item
    {\texttt{cross\_polarized}} (LOGICAL, .false.) By default, the two laser pulses are both linearly polarized in the $z$-direction. If {\texttt{cross\_polarized}} is set to true, then the secondary pulse is linearly polarized in the $y$-direction instead. Note that this overrides the settings for {\texttt{ellipticity}} and {\texttt{ellipticity\_XUV}}.
\end{itemize}

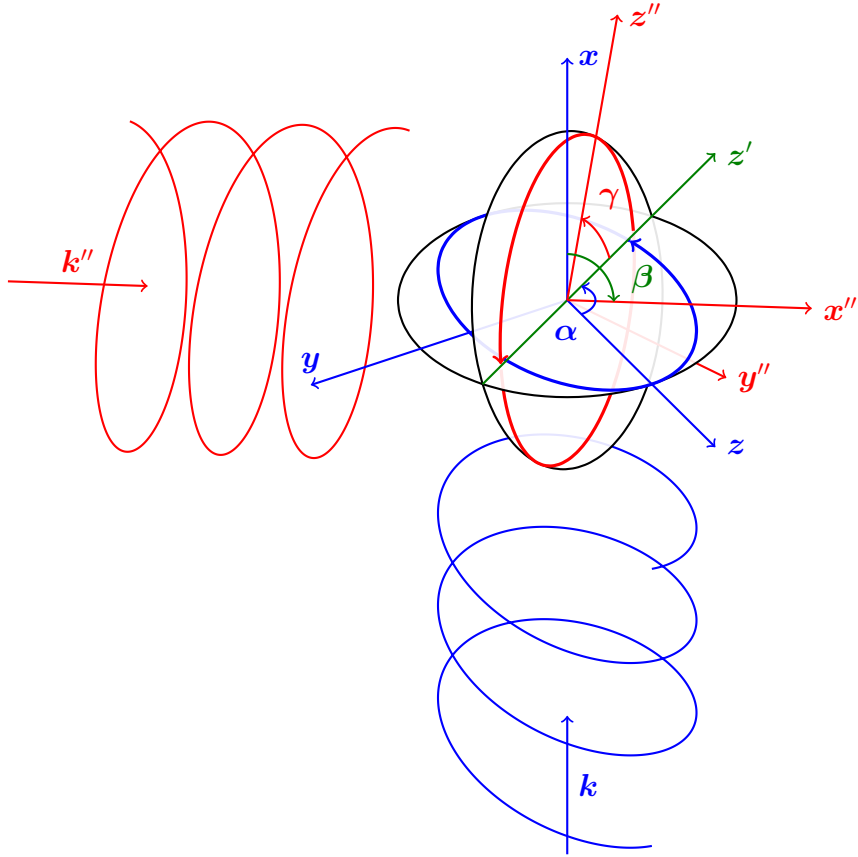
\begin{figure}[htbp]
    \centering
    \tdplotsetmaincoords{55}{120}

\begin{tikzpicture}[tdplot_main_coords,scale=0.75]

    % (real tikz coordinate system: x out, y right, z up)

    % plane radius
    \pgfmathsetmacro{\r}{3}

    % axes scale
    \pgfmathsetmacro{\x}{1.75}

    % major/minor semi-axis lengths
    \pgfmathsetmacro{\a}{3}
    \pgfmathsetmacro{\b}{2}

    % Euler angle arcs
    \pgfmathsetmacro{\e}{0.5}

    % stretch factor for the helixes
    \pgfmathsetmacro{\s}{2.0}

    % left-incoming helix coordinate system (orthonormal right-handed)
    \pgfmathsetmacro{\uax}{-0.707} \pgfmathsetmacro{\uay}{-0.212} \pgfmathsetmacro{\uaz}{ 0.675}
    \pgfmathsetmacro{\ubx}{ 0.707} \pgfmathsetmacro{\uby}{-0.212} \pgfmathsetmacro{\ubz}{ 0.675}
    \pgfmathsetmacro{\ukx}{ 0.000} \pgfmathsetmacro{\uky}{ 0.954} \pgfmathsetmacro{\ukz}{ 0.300}

    % bottom-incoming helix coordinate system (orthonormal right-handed)
    \pgfmathsetmacro{\vax}{ 0.500} \pgfmathsetmacro{\vay}{ 0.866} \pgfmathsetmacro{\vaz}{ 0.000}
    \pgfmathsetmacro{\vbx}{-0.866} \pgfmathsetmacro{\vby}{ 0.500} \pgfmathsetmacro{\vbz}{ 0.000}
    \pgfmathsetmacro{\vkx}{ 0.000} \pgfmathsetmacro{\vky}{ 0.000} \pgfmathsetmacro{\vkz}{ 1.000}

    % left-incoming helix
    \draw[red,thick] plot[smooth,variable=\t,domain=-1800:-720,samples=360] (
        {\uax*\a*cos(\t)+\ubx*\b*sin(-\t)+\ukx*\t*\s/360},
        {\uay*\a*cos(\t)+\uby*\b*sin(-\t)+\uky*\t*\s/360},
        {\uaz*\a*cos(\t)+\ubz*\b*sin(-\t)+\ukz*\t*\s/360}
    );
    \draw[->,red,thick] (-12*\ukx,-12*\uky,-12*\ukz) -- node[above] {$\bm{k}''$} (-9*\ukx,-9*\uky,-9*\ukz);

    % bottom-incoming helix
    \draw[blue,thick] plot[smooth,variable=\t,domain=-1800:-720,samples=360] (
        {\vax*\a*cos(\t)+\vbx*\b*sin(-\t)+\vkx*\t*\s/360},
        {\vay*\a*cos(\t)+\vby*\b*sin(-\t)+\vky*\t*\s/360},
        {\vaz*\a*cos(\t)+\vbz*\b*sin(-\t)+\vkz*\t*\s/360}
    );
    \draw[->,blue,thick] (-12*\vkx,-12*\vky,-12*\vkz) -- node[right] {$\bm{k}$} (-9*\vkx,-9*\vky,-9*\vkz);

    % horizontal plane (left part)
    \draw[thick,fill=white,fill opacity=0.9] plot[smooth,variable=\t,domain=90:270] (
        {-\r*sin(\t)},
        {\r*cos(\t)},
        {0}
    );

    % tilted plane (bottom part)
    \draw[thick,fill=white,fill opacity=0.9] plot[smooth,variable=\t,domain=45:225] (
        {\uax*\r*cos(\t) + \ubx*\r*sin(-\t)},
        {\uay*\r*cos(\t) + \uby*\r*sin(-\t)},
        {\uaz*\r*cos(\t) + \ubz*\r*sin(-\t)}
    );

    % projection of left helix (bottom part)
    \draw[red,very thick] plot[smooth,variable=\t,domain=55:235] (
        {\uax*\a*cos(\t) + \ubx*\b*sin(-\t)},
        {\uay*\a*cos(\t) + \uby*\b*sin(-\t)},
        {\uaz*\a*cos(\t) + \ubz*\b*sin(-\t)}
    );

    % axis y''
    \draw[->,red,thick] (0,0,0) -- (-\x*\ubx*\r,-\x*\uby*\r,-\x*\ubz*\r) node[right] {$\bm{y}''$};

    % horizontal plane (right part)
    \draw[thick,fill=white,fill opacity=0.9] plot[smooth,variable=\t,domain=-90:90] (
        {-\r*sin(\t)},
        {\r*cos(\t)},
        {0}
    );

    % projection of bottom helix  (left part)
    \draw[blue,very thick] plot[smooth,variable=\t,domain=70:250] (
        {\vax*\a*cos(\t) + \vbx*\b*sin(-\t)},
        {\vay*\a*cos(\t) + \vby*\b*sin(-\t)},
        {\vaz*\a*cos(\t) + \vbz*\b*sin(-\t)}
    );

    % axis y
    \draw[->,blue,thick] (0,0,0) -- (-\x*\vbx*\r,-\x*\vby*\r,-\x*\vbz*\r) node[above] {$\bm{y}$};

    % tilted plane (top part)
    \draw[thick,fill=white,fill opacity=0.9] plot[smooth,variable=\t,domain=-135:45] (
        {\uax*\r*cos(\t) + \ubx*\r*sin(-\t)},
        {\uay*\r*cos(\t) + \uby*\r*sin(-\t)},
        {\uaz*\r*cos(\t) + \ubz*\r*sin(-\t)}
    );

    % projection of left helix (top part)
    \draw[<-,red,very thick] plot[smooth,variable=\t,domain=-123:55] (
        {\uax*\a*cos(\t) + \ubx*\b*sin(-\t)},
        {\uay*\a*cos(\t) + \uby*\b*sin(-\t)},
        {\uaz*\a*cos(\t) + \ubz*\b*sin(-\t)}
    );

    % projection of bottom helix  (right part)
    \draw[<-,blue,very thick] plot[smooth,variable=\t,domain=-110:70] (
        {\vax*\a*cos(\t) + \vbx*\b*sin(-\t)},
        {\vay*\a*cos(\t) + \vby*\b*sin(-\t)},
        {\vaz*\a*cos(\t) + \vbz*\b*sin(-\t)}
    );

    % axis z'
    \draw[->,black!50!green,thick] (\r,0,0) -- (-\x*\r,0,0) node[right] {$\bm{z}'$};

    % axis x
    \draw[->,blue,thick] (0,0,0) -- (\x*\vkx*\r,\x*\vky*\r,\x*\vkz*\r) node[right] {$\bm{x}$};

    % axis z
    \draw[->,blue,thick] (0,0,0) -- (\x*\vax*\r,\x*\vay*\r,\x*\vaz*\r) node[right] {$\bm{z}$};

    % axis x''
    \draw[->,red,thick] (0,0,0) -- (\x*\ukx*\r,\x*\uky*\r,\x*\ukz*\r) node[right] {$\bm{x}''$};

    % axis z''
    \draw[->,red,thick] (0,0,0) -- (\x*\uax*\r,\x*\uay*\r,\x*\uaz*\r) node[right] {$\bm{z}''$};

    % Euler alpha
    \draw[<-,blue,thick] plot[smooth,variable=\t,domain=-120:0] (
        {\vax*\e*cos(\t) + \vbx*\e*sin(-\t)},
        {\vay*\e*cos(\t) + \vby*\e*sin(-\t)},
        {\vaz*\e*cos(\t) + \vbz*\e*sin(-\t)}
    ) node[left,below,xshift=-2mm] {$\bm{\alpha}$};

    % Euler beta
    \draw[<-,black!50!green,thick] plot[smooth,variable=\t,domain=-90:0] (
        {\vkx*2*\e*cos(\t) + \ukx*2*\e*sin(-\t)},
        {\vky*2*\e*cos(\t) + \uky*2*\e*sin(-\t)},
        {\vkz*2*\e*cos(\t) + \ukz*2*\e*sin(-\t)}
    ) node[right,below,xshift=1cm] {$\bm{\beta}$};

    % Euler beta
    \draw[<-,red,thick] plot[smooth,variable=\t,domain=0:45] (
        {\uax*3*\e*cos(\t) + \ubx*3*\e*sin(-\t)},
        {\uay*3*\e*cos(\t) + \uby*3*\e*sin(-\t)},
        {\uaz*3*\e*cos(\t) + \ubz*3*\e*sin(-\t)}
    ) node[right,above,yshift=5mm] {$\bm{\gamma}$};

\end{tikzpicture}
    \caption{The polarization plane of the laser pulse used in RMT is designated by setting the three Euler angles, here labelled $\alpha$, $\beta$ and $\gamma$. The default arrangement (polarization in the $yz$-plane) is shown by the blue polarization ellipse and the corresponding blue axes. The red ellipse and axes show the arbitrarily oriented polarization plane, as well as the three Euler angles used for the transformation of that plane.}
    \label{fig:arb_pol_plane}
\end{figure}

The orientation of the coordinate frame is defined by:
\begin{itemize}
    \item 
    {\texttt{euler\_alpha}} (REAL, 0.0)
    \item
    {\texttt{euler\_beta}} (REAL, 0.0)
    \item
    {\texttt{euler\_gamma}} (REAL, 0.0)
\end{itemize}
which are the Euler angles, in degrees, for the orientation of the polarization plane. The $x-z'-x''$ convention is used, i.e., $\alpha$ corresponds to the rotation of the plane around the $x$ axis, $\beta$ corresponds to the rotation around the new $z$ axis, and $\gamma$ corresponds to the rotation around the new-new $x$ axis.  This arrangement is shown in Fig.~\ref{fig:arb_pol_plane}. The possibility to change the orientation of the pulse with respect to the coordinate system is especially useful in molecular calculations. Table~\ref{tab:euler-angles} summarizes some special choices of the Euler angles. 
%The default orientation is such that the polarization ellipse lies in the $yz$ plane, with the major semi-axis directed along the $z$ axis, the minor semi-axis directed along $-y$, and is circulated in accord with the right-hand rule with respect to the positive $x$ direction.

\begin{table}[htbp]
    \centering
    \begin{tabular}{rrrl}
        \toprule
        $\alpha$ & $\beta$ & $\gamma$ & field orientation \\
        \midrule
         0 & 0 & 0 & $\mathbf{E}(t) = \frac{F(t)}{\sqrt{1+\eta^2}} \left(0, -\eta \sin (\omega t + \varphi), \cos (\omega t + \varphi)\right)$ \\
         0 & 0 & $-90$ & $\mathbf{E}(t) = \frac{F(t)}{\sqrt{1+\eta^2}} \left(0, \cos (\omega t + \varphi), \eta \sin (\omega t + \varphi)\right)$ \\ 
         0 & $-90$ & $-90$ & $\mathbf{E}(t) = \frac{F(t)}{\sqrt{1+\eta^2}} (\cos (\omega t + \varphi), 0,\eta \sin (\omega t + \varphi))$ \\
         90 & 90 & 90 & $\mathbf{E}(t) = \frac{F(t)}{\sqrt{1+\eta^2}} (\cos (\omega t + \varphi), \eta \sin (\omega t + \varphi), 0)$ \\
         \midrule
         0 & $\phi - 90$ & $-\theta$ & $\mathbf{E}(t) = F(t) \left(\sin\theta \cos\phi, \sin\theta \sin\phi, \cos\theta \right) \cos (\omega t + \varphi)$ \\
        \bottomrule
    \end{tabular}
    \caption{Selected special combinations of the Euler angles (in degrees) for RMT input. The symbol $\eta$ denotes the ellipticity (numerical eccentricity) as in Eq.~(\ref{def:efield}). The last row of the table is for \(\eta = 0\) only, i.e. linear polarization.}
    \label{tab:euler-angles}
\end{table}

Additionally, the polarization plane can be changed to the $xy$ plane using
\begin{itemize}
    \item {\texttt{xy\_plane\_desired}} (LOGICAL, .false.)
\end{itemize}
This may be desirable, particularly for circularly polarized laser pulses, as it reduces the number of dipole-accessible states relative to the $zy$ plane, where all magnetic sublevels must be retained (see Sec. \ref{sec:arbpol}). The number of outer-region channels is approximately halved compared to calculations adopting the $zy$ plane. Note that this is tantamount to setting the {\texttt{euler\_\{alpha, beta, gamma\}}} angles to 90, and overrides any user-specified values. 

We note that, in principle, there is no reason that the code should be limited
to only two laser pulses, although no options for going beyond this limit are
currently implemented. Should a user wish to implement additional laser pulses
it is reasonably straightforward to modify the {\texttt{electric\_field}} module
to do so: simply use the function {\texttt{E\_Pulse2}} as a template and add a
third (or more as required) call to the {\texttt{pulse\_profile}} function. 

\subsubsection{Calculation Parameters}

\noindent When calculating high-harmonic and absorption spectra, the expectation values of the dipole (length and velocity) operators at each time step are required. These are output using:
\begin{itemize}
    \item 
    {\texttt{dipole\_output\_desired}} (LOGICAL, .false.)
    \item
    {\texttt{dipole\_velocity\_output}} (LOGICAL, .false.)
\end{itemize}
Additionally, in calculations for arbitrary polarization, more than one
component, or perhaps a single component other than the $z$ component, of the expectation value may be desired. In such cases, the calculation of the \(x\), \(y\) and \(z\) components may be activated using
\begin{itemize}
    \item
    {\texttt{dipole\_dimensions\_desired}} (LOGICAL, .false., .false., .true.)
\end{itemize}
which by default enables calculation of the \(z\) component.

\noindent The number of OpenMP threads to employ in the inner and outer regions may be set using:
\begin{itemize}
    \item 
    {\texttt{no\_of\_OMP\_threads\_inner}} (INTEGER, 1)
    \item 
    {\texttt{no\_of\_OMP\_threads\_outer}} (INTEGER, 1)
\end{itemize}
Calculations may be distinguished from others in the same directory by providing a set of characters appended to the file name:
\begin{itemize}
    \item
    {\texttt{version\_root}} (CHARACTER, ``'') The version identifier to append to all output files. The full version suffix will also contain intensity information.
\end{itemize}
Important properties related to the time propagation are:
\begin{itemize}
    \item 
    {\texttt{taylors\_order}} (INTEGER, 8) The order of the time propagator.
    
    \item
    {\texttt{timesteps\_per\_output}} (INTEGER, 20) The interval at which output data is written to file. Importantly, the electric field and dipole data will be output on {\it{every}} time step if {\texttt{dipole\_output\_desired}} is true.
    \item
    {\texttt{checkpts\_per\_run}} (INTEGER, 1) A complete dump of all output data will be performed at every checkpoint, from which the calculation may be restarted.
    \item
    {\texttt{keep\_checkpoints}} (LOGICAL, .false.) Retains all previously output checkpoint data if set to true. This can be used, for instance, to track the evolution of the wavefunction.
\end{itemize}
    A calculation that terminates after a checkpoint (either because it has
    finished, or because it has crashed) may be restarted from the last
    checkpoint. The calculation picks up the restart information from the
    {\texttt{hstat}} output file. The new `initial' state wavefunction is read from the
    {\texttt{state}} directory, and the calculation parameters etc. are read as
    normal from {\texttt{input.conf}}. Note that, in general, the calculation will run for
    an {\it additional} {\texttt{steps\_per\_run\_approx}} time-steps of duration
    $\delta t$ as defined in Eq. \ref{dt_define}. Thus to finish an incomplete
    calculation, these parameters should be redefined to cover only the {\it
    remaining} time.  

    For large-scale calculations, it may be advantageous to limit the number (potentially thousands) of files output by the code. This may be achieved by setting two parameters:
\begin{itemize}
    \item
    {\texttt{binary\_data\_files}} (LOGICAL, .false.) Output the population data as a single, unformatted file. This saves substantial execution time for large calculations when enabled. Individual channel populations can subsequently be recovered using the \texttt{/utilities/py\_lib/data\_recon.py} python script.
    \item
    {\texttt{write\_ground}} (LOGICAL, .true.) Write the ground-state wavefunction to file. When set to \texttt{false} in large calculations, this avoids the writing of many files to the \texttt{ground} directory.
\end{itemize}
In addition to the standard output, detailed information on calculation parameters and arrays may be desirable. This can be obtained using the flag 
\begin{itemize}
    \item
    {\texttt{debug}} (LOGICAL, .false.) Write extra debugging information to the screen at run-time.
\end{itemize}
In the outer region, it is possible to specify whether or not the long-range potentials $W_E$, $W_P$ and $W_D$ should be included in a calculation:
\begin{itemize}
    \item
    {\texttt{wd\_ham\_interactions}} (LOGICAL, .true.) Include long-range interactions between the laser field and residual ion. 
     \item
    {\texttt{we\_ham\_interactions}} (LOGICAL, .true.) Include long-range interactions between the ionized electron and residual ion. 
     \item
    {\texttt{wp\_ham\_interactions}} (LOGICAL, .true.) Include long-range interactions between the laser field and the ejected electron. 
\end{itemize}
For calculations outputting the dipole expectation values, it may be desirable to set:
\begin{itemize}
     \item
    {\texttt{window\_harmonics}} (LOGICAL, .true.) Applies a Gaussian mask to the $W_P$ matrix so that the expectation value of the dipole is not dominated by contributions at large radial distances. The position and shape of the mask are controlled by:
    \item
    {\texttt{window\_cutoff}} (REAL, 100.00) Starting radial co-ordinate of the Gaussian mask applied to $W_P$.
    \item
    {\texttt{window\_FWHM}} (REAL, 50.00) The full-width-half-maximum of the Gaussian mask applied to $W_P$.
\end{itemize}
The code has support for including an absorbing boundary in the outer region:
\begin{figure}[ht]
    \centering
    \includegraphics[width=14cm]{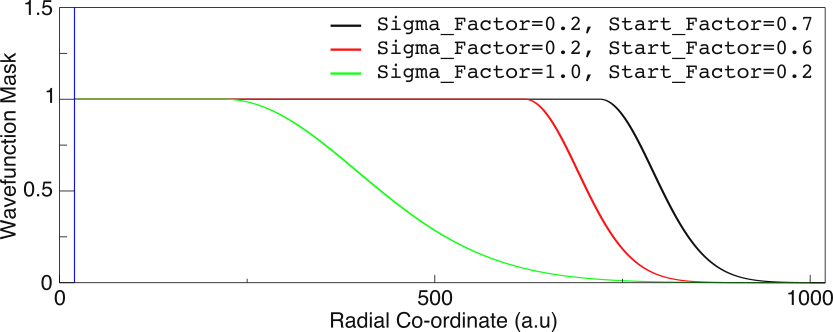}
    \caption{The Gaussian mask applied to the outer region wavefunction when {\texttt{absorb\_desired = .true.}}. {\texttt{start\_factor}} controls what percentage of the outer region is unmasked, and {\texttt{sigma\_factor}} controls the severity of the masking.
    \label{fig:mask_function}}
\end{figure}
\begin{itemize}
    \item 
    {\texttt{absorb\_desired}} (LOGICAL, .false.)
    \item
    {\texttt{start\_factor}} (REAL, 0.7)
    \item
    {\texttt{sigma\_factor}} (REAL, 0.2)
    \item
    {\texttt{absorption\_interval}} (INTEGER, 20)
\end{itemize}
\noindent Essentially, the absorbing boundary is a Gaussian mask which multiplies the outer region wavefunction after every {\texttt{absorption\_interval}} iterations.  The un-masked proportion of the outer region is set by {\texttt{start\_factor}}.  A value of {\texttt{start\_factor}} = 0.7 starts the absorbing boundary 70$\%$ into the outer region;  {\texttt{sigma\_factor}} sets the severity of the absorption. The larger this is, the gentler the absorption. The mask function is defined by 
\begin{equation}
    M(x)  =  \begin{cases} 
    1  \quad   \mbox{ for } x< x_s \\ \\
     \exp\left[-\left(\frac{x-x_s}{\sigma}\right)^2\right]
    \quad  \mbox{ for } x_s \le x \le x_l 
    \end{cases}
\end{equation}
where $x_l$ is the outermost point in the outer region, $x_s = $ {\texttt{start\_factor}} $\times x_l$ and $\sigma = 0.5 \times$ {\texttt{sigma\_factor}} $\times x_l$. The mask function is shown in Fig. \ref{fig:mask_function} for a variety of parameters. Note that the outer region is of size 1000 a.u., but the outermost point has a radial distance of 1020 a.u., since the inner region is 20 a.u. in extent.

\subsection{Convergence Testing}
Although the particulars of any given calculation vary substantially, there are a few checks that should always be performed to ensure convergence in the results. Specifically it should be ensured that
\begin{itemize}
  \item
    increasing the maximum angular momentum retained (RMatrix I, RMatrix II or
    UKRMol+)
  \item
    increasing the size of the R-matrix radius (RMatrix I, RMatrix II or
    UKRMol+)
  \item
    increasing the extent of the outer region
    ({\texttt{no\_of\_pes\_to\_use\_outer}}, {\texttt{x\_last\_master}} and
    {\texttt{x\_last\_others}})
  \item
    decreasing the time-step ({\texttt{steps\_per\_run\_approx}} and
    {\texttt{final\_T}})
  \item 
    decreasing the grid spacing ({\texttt{deltaR}})
\end{itemize}
do not modify the result of interest. We note that even in calculations employing an absorbing boundary, simply modifying the extent of the outer region should reveal any problems with the absorbing boundary parameters.

\section{Parallel Implementation}
\label{rmt-code-computational-considerations}

\subsection{Overview}\label{overview}
The RMT code has been designed for implementation on massively parallel architectures. As such, there is a minimum number of computer cores required for any given calculation. The outer region always requires at least one core. The minimum number of cores for the inner region is set by the number of symmetry blocks in the calculation. In principle, this may be quite small (for single photon processes using linearly polarized light, only two or three cores may be required for instance), and in the molecular case it will always be possible, in principle, to execute the code with $n\leq 8$ cores (for the inner region).% as this is the maximum number of symmetry blocks which may be included for the most symmetric molecular point group, D$_{2h}$.  

For the most general atomic case, the minimum number of inner-region cores is given by
\begin{equation}
     n \geq 2\left[\left(M_{L_{\mbox{max}}} + 1\right)^2 + \left(2\;M_{L_{\mbox{max}}}\; + 1\right)\left(L_{\mbox{max}} - M_{L_{\mbox{max}}}\right)\right],
\end{equation}
where $L_{\mathrm{max}}$ is the highest angular momentum, and $M_{L_{\mathrm{max}}}$ is the highest angular momentum projection included in the calculation. We note the following special cases:

\begin{itemize}
    \item 
the parameter {\texttt{ML\_max}} is not set or is set equal to $L_{\mathrm{max}}$ (this is the default behaviour):
\begin{equation}
    n \geq 2(L_{\mathrm{max}} + 1)^2.
\end{equation}
\item
the parameter {\texttt{ML\_max} = 0} (this should be used for atomic calculations with linearly polarized light oriented along the $z$-axis, which is the default orientation):
\begin{equation}
    n \geq L_{\mathrm{max}} + 1.
\end{equation}
\item
the parameter {\tt xy\_plane\_desired = true} (this should be used for calculations with circular or elliptical polarization):
\begin{equation}
    n \geq (L_{\mathrm{max}} + 1)^2.
    \label{coresxyplane}
\end{equation}
\end{itemize}
In the case of $xy$-plane polarization, the parameter {\texttt{ML\_max}} will be set equal to $L_{\rm max}$, and so no further reduction in the number of cores given by Eq.\;\eqref{coresxyplane} is possible.

In practice, however, realistic calculations will require more cores, even in the molecular case. Should a calculation be executed using fewer than the minimum number of cores, the code will exit with an error message showing the minimum number required. Because of the several layers of parallelism employed in both regions, it is possible to obtain good scaling performance by adding more cores to the inner region, and the code will attempt to assign the cores to balance the load within the inner region. 

All of that said, for very large calculations, the bottleneck will typically be the outer region, where loops over the large number of electron-emission channels included take the longest time. The weak scaling \cite{gustafsons_law} of the outer region is almost perfect: adding more cores to facilitate an increase in the extent of the outer region incurs little cost. In terms of strong scaling \cite{Amdahls_law}, decreasing the number of grid points used on each core can yield a speed up, but some test calculations should always be performed to find the optimal balance of speed and absolute number of cores.

The largest RMT calculations to date employed 16416 cores (10104 inner, 6312 outer) on the UK supercomputing facility, ARCHER. 

\subsection{Parallelization Strategy}\label{parallelization-strategy}

The code employs the standard R-matrix paradigm of dividing
configuration space into two regions. Within each region, a different
parallelization strategy is employed. In any given calculation, the
bottleneck exists in one of these two regions. For calculations
comprising a high degree of atomic or molecular structure, the inner region tends to
dominate. For those comprising many channel functions (large angular
momentum expansions) the outer region can dominate. The skill in
optimizing the calculation is to balance the workload in each region by
a judicious allocation of cores.

\begin{figure}[ht]
\centering
\includegraphics[width=0.7\textwidth]{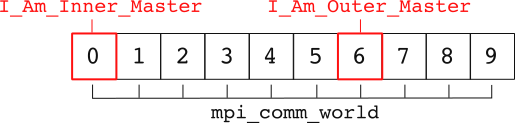}
\caption{The first MPI task in each of the inner and outer regions is termed the region master. All communication between the regions takes place between these two cores \label{fig:masters}}
\end{figure}

Communication between the two regions is handled by the `region-master'
cores (essentially the first core in each region , see Fig. \ref{fig:masters}). Rather than having a
separate communicator for this, every core has a logical flag set at the
start of the calculation: \texttt{i\_am\_inner\_master} is set on all
inner region cores and is true only for the inner-region master.
Similarly, \texttt{i\_am\_outer\_master} is set for the outer region
cores. Subroutines named
{\texttt{first\_PEs\_share\_<...>}} or
\texttt{first\_PE\_receives\_<...>} are called from all cores in a given region, and the logical
flags are used to determine which cores are involved in the
communication using \texttt{MPI\_Comm\_World}.

\subsubsection{Inner-region
parallelization}\label{inner-region-parallelization}

The calculation in the inner region involves repeated
matrix-vector multiplications. This calculation is parallelized in three
layers using both distributed (MPI) and shared (OpenMP) paradigms.

\paragraph{Layer 1}\label{layer-1}

\begin{figure}[ht]
\centering
\includegraphics[width=0.7\textwidth]{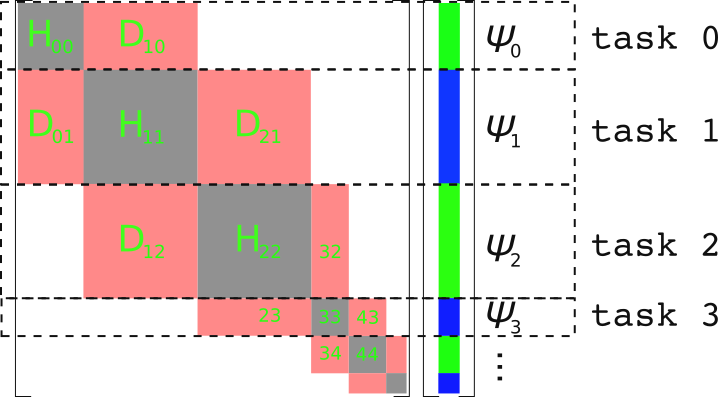}
\caption{Layer 1 parallelism in the inner region. The Hamiltonian matrix and wavefunction vector are divided into symmetry blocks, with each block assigned to (at least one) MPI task. The diagonal block $H_{ii}$ consists of the energies of the states conforming to symmetry $i$, while the off-diagonal block $D_{ij}$ expresses the dipole coupling between states in blocks $i$ and $j$. \label{fig:inner_parallelism}}
\end{figure}

Both the Hamiltonian matrix and the wavefunction vector are divided
into symmetry blocks. As an example, for atomic calculations with a linearly polarized pulse, these symmetry blocks correspond to states of a given angular momentum, as
shown in Fig. \ref{fig:inner_parallelism}. The first layer of parallelism entails the assignment of each symmetry block to an MPI task. In the simplest arrangement, one MPI
task is assigned to each block, so (referring to Fig. \ref{fig:inner_parallelism})
$H_{00}$, $D_{10}$ and $\psi_{0}$ are local to MPI
task 0, $H_{11}$, $D_{01}$, $D_{21}$ and
$\psi_{1}$ are local to MPI task 1, etc.. Here, $H_{ii}$ denotes the diagonal block comprising the energies of the states in symmetry block $i$,
while $D_{ij}$ consists of the dipole matrix elements
coupling states in block $i$ to states in block $j$.

\begin{figure}[t]
\centering
\includegraphics[width=0.7\textwidth]{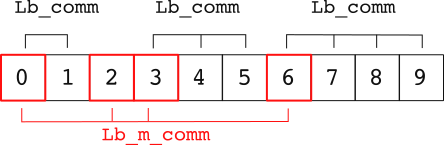}
\caption{Communication strategy for the inner region. Each symmetry block is controlled by a block master. All cores assigned to a symmetry block communicate within the MPI communicator \texttt{Lb\_comm}, and communication between block masters happens using the MPI communicator \texttt{Lb\_m\_comm}.\label{fig:block_parallelism}}
\end{figure}

The communication for this layer is handled using an array of coupling values which, for each pair of MPI task IDs, stores the status of the coupling. Each MPI task then stores a list of the tasks with which it will need to exchange data.
At the beginning of each iteration, the data is transferred
between the tasks using the communicator \texttt{Lb\_m\_comm} (Lb for L
block, m for master, see Fig. \ref{fig:block_parallelism}). This occurs in the subroutine
\texttt{parallel\_matrix\_vector\_multiply\_zm()} in the module
\texttt{live\_communications}.

\paragraph{Layer 2}\label{layer-2}

Evidently, there are systems where certain symmetries will include
substantially more states than others. In such a case, and as suggested by Fig.
\ref{fig:inner_parallelism}, the dipole blocks can also be substantially
larger, and the number of multiplications required on each iteration grows with
the square of the size of the block. Hence the code has the flexibility to
assign multiple MPI tasks to each block. The tasks are allocated by a routine
which first assigns one task per block, and then calculates which block has the
most work per task, assigning an additional task until all have been allocated.
Table \ref{tab:task_allocation} shows a typical allocation of 144 tasks to the 10 blocks used
in a model calculation for Ne$^+$.

\begin{table}[ht]
\caption{A typical allocation of 144 tasks among 10 symmetry blocks in a model calculation for Ne$^+$. \label{tab:task_allocation}}
\centering
\begin{tabular}{lll}
\hline
\hline
Block & States & Tasks\tabularnewline
\hline
0 & 429 & 7\tabularnewline
1 & 970 & 29\tabularnewline
2 & 1126 & 37\tabularnewline
3 & 979 & 29\tabularnewline
4 & 726 & 17\tabularnewline
5 & 500 & 9\tabularnewline
6 & 369 & 5\tabularnewline
7 & 305 & 4\tabularnewline
8 & 284 & 4\tabularnewline
9 & 276 & 3\tabularnewline
\hline
\end{tabular}
\end{table}

Each block is now controlled by the so-called `block master', which
communicates with other block masters and distributes data to the
tasks within the block. Each task within a block handles an equal number
of rows of the matrix-vector multiplication. The block master distributes the relevant portions of
the dipole blocks at the start of the
calculation, as well as of the wavefunction on each iteration.
Communication within each block is handled on the communicator
\texttt{Lb\_comm}.

\paragraph{Layer 3}\label{layer-3}

The final layer of parallelism in the inner region is shared-memory
parallelization on each MPI task. From layers 1 and 2, each MPI task has
a chunk of the Hamiltonian matrix and wavefunction vector with which it
performs matrix-vector multiplications. The shared-memory parallelism is
implemented in two ways.

First, several do-loop structures with independent loops are farmed out
to the shared memory threads with simple \texttt{!\$OMP~PARALLEL~DO} sentinels.
Secondly, all library routines for linear algebra that support
shared-memory processing will execute in parallel on all available
shared memory threads.

\subsubsection{Outer-region
parallelization}\label{outer-region-parallelization}

The outer-region parallelization is somewhat easier to envisage than that for the
inner region, as it is actually a division of physical space. Two layers
of parallelism are employed in the outer region, one using MPI and one
using OpenMP.

\paragraph{Layer 1}\label{layer-1-1}

The major division in the outer region is one of the physical
space itself. Thus, an outer region of 100 a.u. might be divided into four smaller
sectors of 25 a.u., with each sector handled by an MPI task. Because the
outer region uses an explicit, grid-based representation of the
wavefunction, this corresponds to each MPI task handling a set number of
grid points.

\begin{figure}[ht]
\centering
\includegraphics[width=0.7\textwidth]{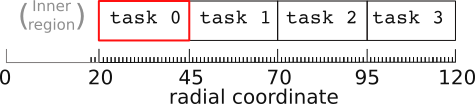}
\caption{Layer 1 parallelization in the outer region. Each outer-region MPI task handles a subset of the entire physical space. Communication is only required between nearest neighbours.}
\label{fig:outer_grid}
\end{figure}

In practice, the outer-region master (first MPI task in the outer
region, highlighted red in Fig. \ref{fig:outer_grid}) is allocated a smaller number of grid
points than the rest of the outer-region tasks, freeing additional resource
for the extra communication responsibilities with the inner region.

Performance is enhanced by reducing the number of grid points per sector
(and increasing the number of outer-region MPI tasks to maintain the
size of the outer region). However, the finite-difference rule
implemented requires a minimum of $2 \times$ \texttt{taylors\_order} grid points ($2\times8= 16$ by default) per sector. If this
limit is reached, further performance can be extracted from the second
layer of parallelism.

\paragraph{Layer 2}\label{layer-2-1}

As in the inner region, additional performance can be obtained with the
use of OpenMP parallelism of each MPI task. Thus, a number of shared-memory threads can be allocated per outer-region task (this is
controlled separately from the number of OpenMP threads in the inner
region). In the outer region, most of the calculation takes place for
each electron-emission channel independently of the others, so all do-loop structures which loop over the variable
\texttt{channel\_ID} are parallelized with the
\texttt{!\$OMP~PARALLEL~DO} structure.

\section{Output files}
\subsection{List of output files}\label{list-of-output-files}

The following files are updated during the calculation every
\texttt{timesteps\_per\_output} iterations:

\begin{itemize}
\item
  \texttt{CurrentPosition} Summary of the current status of the calculation (key parameters and values of variables).
\item
  \texttt{pop\_all.\textless{}version\_number\textgreater{}} Total population (should be normalised to 1.0).
\item
  \texttt{pop\_inn.\textless{}version\_number\textgreater{}} Total population in the inner region.
\item
  \texttt{pop\_out.\textless{}version\_number\textgreater{}} Total population in the outer region. Can be used as a proxy for ionization yield.
\item  
   \texttt{EField.\textless{}version\_number\textgreater{}}  The electric field strength (in a.u.) in component form (separate columns for the $x,y,z$ components).
\end{itemize}

\noindent
If {\texttt{timings\_desired == .true.}} then the following timing files are updated at every time step:
\begin{itemize}
\item
  \texttt{timing\_inner.\textless{}version\_number\textgreater{}} Timing information recorded on the inner-region master.
\item
  \texttt{timing\_outer0.\textless{}version\_number\textgreater{}} Timing information recorded on the outer-region master.
\item
  \texttt{timing\_outer1.\textless{}version\_number\textgreater{}} Timing information recorded on the first (non-master) outer-region core. The two outer-region timing files allow the balancing of work on the outer-region master.
\end{itemize}

\noindent
Additionally:
\begin{itemize}
\item
  \texttt{hstat.\textless{}version\_number\textgreater{}} Checkpoint status file which is read by the RMT code for restart. Written only at each checkpoint.
\end{itemize}

\subsection{Optional outputs}\label{optional-outputs}

\begin{itemize}
\item
  \texttt{expec\_z\_all.\textless{}version\_number\textgreater{}}
  \texttt{(dipole\_output\_desired=.true.)}
    The expectation value of the dipole length operator at every time step in component form (separate columns for the \(x\), \(y\), and \(z\) components). 
\item
  \texttt{expec\_v\_all.\textless{}version\_number\textgreater{}}
  \texttt{(dipole\_velocity\_output=.true.)} and
  \texttt{(dipole\_output\_desired=.true.)}
    The expectation value of the dipole velocity operator at every time step in component form (separate columns for the \(x\), \(y\), and \(z\) components).
\end{itemize}

Note that if \texttt{dipole\_output\_desired=true}, then the electric
field is also output at every time step. This can be necessary to obtain
high-resolution, high-energy Fourier Transforms of the output data.

\subsection{Channel populations}\label{channel-populations}

The \texttt{data} directory contains plain text files describing the population after every \texttt{timesteps\_per\_output} iterations in

\begin{itemize}
\item
  the ground state, and
\item
  each of the electron emission channels.
\end{itemize}

The channel numbering follows the standard R-matrix protocol. In the atomic case, it is sorted in ascending order by

\begin{enumerate}
\def\labelenumi{\arabic{enumi}.}
\item
  the total angular momentum of the final state,
\item
  the parity of the final state (even then odd),
\item
  the magnetic sublevel $(M_L)$ of the final state,
\item
  the energy of the residual-ion state to which the emitted electron is
  coupled, and
\item
  the angular momentum of the emitted electron.
\end{enumerate}

In the molecular case, the channel number is sorted by:

\begin{enumerate}
\def\labelenumi{\arabic{enumi}.}
\item
  the irreducible representation to which the final state belongs,
\item
  the energy of the residual-ion state to which the emitted electron is
  coupled,
\item
  the angular momentum of the emitted electron, and
\item
  the projection of the emitted-electron angular momentum along the $z$ axis ($m_l$).
\end{enumerate}

Note that if the input parameter \texttt{binary\_data\_files = .true.}, then the
channel population files are not written in plain text, but as a single, binary
file using stream-io. This can yield substantial run-time savings when many
outer-region channels are included in the calculation. A python utility for
reconstructing the population files in post-processing is provided:
\texttt{/utilities/py\_lib/data\_recon.py}. 

\subsection{Wavefunction data}\label{sec:wavefunction-data}

The initial and final wavefunctions are recorded in the \texttt{ground} (provided \texttt{write\_ground} is not set to \texttt{.false.} in \texttt{input.conf})
and \texttt{state} directories respectively. The wavefunction is written
out in parallel by each MPI task, and must be
reconstructed from the binary files in post-processing. A utility
code (\texttt{reform}) is provided to accomplish this, the details of which can be found in Sec. \ref{sec:reform}.

\section{Observables}\label{sec:observables}

%wavefunction: how processed for electron energy/momentum plots
%dipole: how processed for HHG/ATAS 

Much recent interest in ultrafast physics stems from enhanced experimental capabilities harnessing spectroscopic accuracy to resolve the details of electronic structure and dynamics \cite{attosecond_spectroscopy_review}. RMT affords the computation of photo-absorption and emission spectra directly from the laser-induced dipole moment of the atomic or molecular system.

\subsection{Photoemission (High-harmonic generation)}
Fundamentally, light is produced by accelerated charges. In the cases considered by RMT, it is the laser-induced dipole moment of the atomic or molecular system which gives rise to photoemission. Following Ref. \cite{brown_helium}, we may show that the electric field produced by an oscillating dipole is

%Fundamentally, light is produced by accelerated charges, and in the cases considered by RMT, the charge in question is the dipole. Following Ref. \cite{brown_helium}, we may show that the electric field produced by an oscillating dipole is 
%
\begin{equation}
  \mathbf{E}(t) \propto \mathbf{\ddot{d}}(t) =
  \frac{d^2}{dt^2}\langle\Psi(t)|\mathbf{D}|\Psi(t)\rangle,
\end{equation}
where $\mathbf{\ddot{d}}(t)$ is the time-dependent expectation value of the dipole acceleration, $\mathbf{D}$ is the dipole operator, and $\Psi(t)$ is the wavefunction. Then, the power spectrum of the emitted radiation is given, up to a proportionality constant,  by $|\mathbf{\ddot{d}}(\omega)|^2$, the Fourier transform of $\mathbf{\ddot{d}}(t)$ squared. The dipole acceleration $\mathbf{\ddot{d}}$ cannot, however, be computed so easily (except in simple cases such as atomic helium \cite{brown_helium}), as this quantity is prohibitively sensitive to the description of atomic structure at very small radial distances. Instead, the relationships between acceleration, velocity and displacement can be exploited to express the harmonic spectrum in terms of the dipole velocity and/or length.

Thus, in RMT calculations, the harmonic spectrum is calculated either in the length form, using the time-dependent expectation value of the dipole
operator $\mathbf{D}$,
\begin{equation} \mathbf{d} (t) =  \langle \Psi(t) | \mathbf{D} | \Psi(t)\rangle
  \notag, \end{equation}
 or in the velocity form, using the dipole velocity operator $\mathbf{\dot{D}}$,
\begin{equation} \dot{\mathbf{d}} (t) =  \langle \Psi(t) | \mathbf{\dot{D}} | \Psi(t)\rangle
  \notag. \end{equation}
For each form, RMT calculates all three Cartesian components of the expectation value.
The harmonic spectrum is then given by
\begin{equation} S(\omega) \quad = \quad \omega^4 | \mathbf{d}(\omega)|^2
 \quad= \quad \omega^2 | \mathbf{\dot{d}}(\omega)|^2
  \notag,
\end{equation}
where $\omega$ is the photon energy, and $\mathbf{d}(\omega)$ and
$\mathbf{\dot{d}}(\omega)$ are the Fourier
transforms of $\mathbf{d}(t)$, and of $\mathbf{\dot{d}}(t)$, respectively. Consistency between the spectra in length
and velocity form constitutes an important test of accuracy for RMT
calculations.

A python script is provided with the code, under
{\texttt{/utilities/py\_lib/gen\_hhg.py}}, to compute the length and velocity forms of the harmonic spectrum.  We note that currently, for molecular calculations, only the length form spectra are computed.

\subsection{Photoabsorption (Transient absorption Spectroscopy)}

Following Refs. \cite{gaarde_absorption_theory,ATAS_theory_2}, it can be shown that the transient absorption spectrum $\sigma(\omega)$ is given by
\begin{equation}
\sigma (\omega) = 4\pi\alpha\omega \; \textnormal{Im} \bigg[\frac{\mathbf{d}(\omega)}{\mathbf{E}(\omega)}\bigg],
\end{equation}
where $\alpha$ is the fine-structure constant, $\omega$ is the photon energy, and $\mathbf{d}(\omega)$ and $\mathbf{E}(\omega)$ are the Fourier transforms of the time-dependent expectation value of the dipole operator $\mathbf{d}(t)$, and of the electric field of the driving laser $\mathbf{E}(t)$, respectively.

The transient absorption spectrum can therefore be calculated from the time-dependent dipole expectation, generated in RMT as described above, and the electric field. 

A python script is provided with the code, under
{\texttt{/utilities/py\_lib/gen\_tas.py}}, to compute the absorption spectrum.

%channel pops
\subsection{Photoelectron spectra}\label{sec:reform}

Photoelectron energy and momentum spectra can be generated using the wavefunction data output into the directories {\texttt{/ground}} (initial) and {\texttt{state}} (final).  The utility code {\texttt{reform}} is available in the main {\texttt{/source}} directory, and is compiled against the same common code base as RMT. The code must be run from either the \texttt{ground} or \texttt{state} directories, and thus looks in the parent directory (whence the \texttt{rmt.x} executable is run) for input files. This is important if performing post-processing on a different machine, where soft links to input files may not be preserved.

\subsubsection{Inner region wavefunction}\label{inner-region-wavefunction}

Each inner-region block master writes out the spline coefficients
for a set of channel wavefunctions in a file
\texttt{psi\_inner\textless{}block\_ID\textgreater{}.\textless{}version\_number\textgreater{}}.
To reconstruct the wavefunction in each channel, the utility code reads
the spline information from the input files \texttt{Splinedata} and
\texttt{Splinewaves}, and the atomic-structure information from the
input file \texttt{H}.

The outputs of the utility script are the files
\texttt{InnerWave\textless{}Channel\_ID\textgreater{}}, where the channel
ID is as described above, containing the value of the wavefunction at
a set of inner-region grid points. The grid spacing is set to the value of \texttt{deltaR} from \texttt{input.conf}. Note that inner-region wavefunctions
are produced only in atomic mode (i.e., not for molecular runs).

\subsubsection{Outer region
wavefunction}\label{outer-region-wavefunction}

Each outer-region MPI task writes out its wavefunction data per channel to a file named
\texttt{psi\_outer\textless{}outer\_ID\textgreater{}.\textless{}version\_number\textgreater{}},
where \texttt{\textless{}outer\_id\textgreater{}} is the
rank of the outer-region MPI task labelled from \texttt{0} to
\texttt{no\_of\_PEs\_to\_use\_outer-1}. The utility script reads in the
data for each \texttt{\textless{}outer\_ID\textgreater{}}, calculates
what the corresponding grid points should be, and writes out the
wavefunction for each channel in files
\texttt{OuterWave\textless{}Channel\_ID\textgreater{}}, where the channel
ID is as described above.

\subsubsection{Execution}\label{execution}

\texttt{reform} should be run from the \texttt{ground} or \texttt{state} directory: \texttt{<path\_to\_compile\_directory>/bin/reform}.
The code does have OpenMP loop directives enabled, so multicore machines
can be exploited simply by exporting the variable
\texttt{\$OMP\_NUM\_THREADS} prior to execution.

There are several
command-line options that can be used to control further processing of the outer-region wavefunction data by \texttt{reform}.
When the \texttt{-{}-density} switch is used, \texttt{reform} will also write the
outer-region position density distribution \(\rho(\bm{r})\) to a file.
In atomic calculations, the position density is given by 

\begin{equation}
    \rho(\bm{r}) = \sum_{n L_n M_n} \left | \sum_{\ell m L M_L} 
    (L_n M_n \ell m|L M_L)
    i^\ell F_{L_n M_n \ell m  L  M_L}(r) {\mathcal{Y}}_{\ell}^{m}(\hat{\bm{r}}) \right |^2 .
    \label{momdiseq}
\end{equation}
Here \(n\) is the index of a residual-ion state with total orbital angular momentum \(L_n\), and \(M_n\) is a projection of that angular momentum. $L$ and $\ell$ are the total angular momenta of the \((N+1)\)-electron system and the outgoing electron respectively, and $M_L$ and $m$ are their projections. \((L_n M_n \ell m|L M_L)\) is a Clebsch-Gordan coefficient, ${\mathcal{Y}}_{\ell} ^m$ is a complex spherical harmonic conforming to the Fano-Racah phase convention, and \(F_p(r)\) is the reduced radial wavefunction of the ejected electron moving in the channel with quantum numbers $p$ \cite{tdrm}. For the molecular case, we have instead
\begin{equation}
    \rho(\bm{r}) = \sum_n \left | \sum_{\ell m} F_{n, \ell m}(r) X_{l,m}(\hat{\bm{r}}) \right |^2,
    \label{molecule_density}
\end{equation}
where \(n\) is the index of the residual-ion state and, at variance with the atomic case, the $X_{l,m}(\hat{\bm{r}})$ are {\it real} spherical harmonics (see~\ref{conventions}).
For convenience of notation, we have omitted all spin quantum numbers. 

With the \texttt{-{}-momentum} switch, \texttt{reform} will produce data for the momentum
density distribution \(\rho(\bm{k})\), an example of which is shown in Fig.~\ref{ar8ev}. 
The momentum densities use the same formulae given in Eqns. (\ref{momdiseq}) and
(\ref{molecule_density}), with the ejected-electron radial wavefunctions \(F_p(r)\) replaced by their Fourier transforms \(\tilde{F}_p(k)\).
As a check of accuracy, the code calculates the channel populations in both position and momentum spaces, aiming to verify norm conservation by the Fourier transform, and to recover the total outer-region population upon summation of all channel populations.
Note that the normalization convention is such
that to recover the total outer-region population as printed by RMT, one would need
to integrate the distributions \(\rho(\bm{r})\) and
\(\rho(\bm{k})\) as \(\int r^{-2} \rho(\bm{r}) \mathrm{d}^3 \bm{r}\) and
\(\int k^{-2} \rho(\bm{k}) \mathrm{d}^3 \bm{k}\), respectively.

The position and momentum density output files, \texttt{OuterWave\textunderscore density.txt} and
\texttt{OuterWave\textunderscore momentum.txt} respectively, have the form
of plain text files. They contain
matrices of densities evaluated per angle (rows) and per radial point (columns)
in the chosen sampling plane, which defaults to the polarization plane, and where the angle is described in the same sense as the time-dependent polarization vector. The spacing between the radial samples in the position density file is equal to the
parameter \texttt{deltaR} from the RMT input namelist, and the first sample corresponds to
the density value at the R-matrix boundary. The spacing between the radial samples in the
momentum density file is equal to \(\Delta k = 2\pi / (R_{\mathrm{max}} - R_{\mathrm{min}})\),
where \(R_{\mathrm{max}}\) is the extent of the outer-region radial grid, and \(R_{\mathrm{min}}\)
is the smallest radius used in the Fourier transform.

Other command-line options are
\begin{itemize}
    \item \texttt{-{}-ntheta} \textit{N},
          default 360: set the number of angular samples around the full circle.
          Default is one-degree resolution in both the polar and azimuthal angles.
    \item \texttt{-{}-nmaxpt} \textit{N}, default 10000: set the maximum number
      of radial samples
          (limited by the actual number of finite-difference grid points in the outer
          region).
    \item \texttt{-{}-channels} \textit{i,j,k}$\ldots$: select only specific channels to include in processing.
    \item \texttt{-{}-rskip} \textit{R}, default 200 (a.u.): ignore part of the outer-region wavefunction closest to the R-matrix boundary
          when evaluating the momentum distribution. This screens out high-lying
          Rydberg states that would obscure the results.
    \item \texttt{-{}-plane} \(\alpha,\beta,\gamma\): choose a different orientation of the sampling plane.
    \item \texttt{-{}-vtk}: produce also a fully three-dimensional distribution in the Visualization Toolkit
          unstructured grid format.
\end{itemize}

The text files contain one more line in addition to the number given by \texttt{-{}-ntheta};
the first line is repeated at the end to allow continuous plotting. The program also writes
one extra output file for each residual-ion state (and for each of its magnetic sub-levels in atomic
cases), with the partial photoelectron distribution corresponding to ionization into that particular state.

\section{Test Suite}
A test suite is provided with sample outputs under \texttt{/tests}. The exemplar calculations are subdivided into atomic and molecular categories, with the atomic calculations further divided into small (can be run on fewer than 96 cores) and large (requires more than 196 cores) directories.

Each atomic test-calculation directory is structured identically. Taking \texttt{tests/atomic\_tests/small\_tests/helium} as an example, the directory contains subdirectories \texttt{inputs} (with the necessary input files \texttt{H}, \texttt{d}, \texttt{d00}, \texttt{Splinedata}, \texttt{Splinewaves} and \texttt{input.conf}), and \texttt{rmt\_output} (including the output files \texttt{pop\_all}, \texttt{pop\_inn}, \texttt{pop\_out}, \texttt{expec\_z\_all} and \texttt{expec\_v\_all}).

The molecular test-calculation directories contain the input files not only for RMT, but also for the UKRmol+ calculations (which generate the input for RMT) in the directory \texttt{UKRmol+}. The input for the RMT calculation is contained in \texttt{/inputs},  where files analogous to the atomic case reside: input files (\texttt{input.conf, molecular\_data}) and output files in the directory \texttt{rmt\_output}. The molecular calculations all run on just 10 cores.

It should not be expected that the output files contained in the
{\texttt{rmt\_output}} directories should be
reproduced identically on all systems with all compilers. It is sufficient to
check that the numbers do not vary substantially. For reference, the data provided was determined using the Intel
17 compiler on ARCHER (www.archer.ac.uk). Furthermore, only some of the
output files produced by the RMT calculation are present in the
{\texttt{rmt\_output}} directories, and thus the utility script
{\texttt{compare\_rmt\_runs}} will show a warning when executed against the
test calculations. 

\section{Compilation}
The RMT code can be obtained from the RMT repository \cite{RMT_repo}. 
Compilation of the code requires CMake version 3.0 or higher, the LAPACK
and MPI libraries, as well as a parallel Fortran compiler. The code has been tested with
Intel (versions 12 -- 19), gfortran and Cray compilers.

CMake automatically interrogates the system to determine the location of
libraries, so depending on system architecture, it may be necessary to export
environment variable \texttt{LD\_LIBRARY\_PATH} to point to the location of the
LAPACK or MKL libraries. Additionally, CMake uses the environment variable
\texttt{FC} as the fortran compiler, thus it may need to be set as

\texttt{>> export FC=\$(which mpif90)}\\

To compile, we recommend creating a separate build
directory, for example in the root directory of the repository:\\

\texttt{>> cd <path to repository>/rmt}

\texttt{>> mkdir build}

\texttt{>> cd build}

\texttt{>> cmake ../source}

\texttt{>> make}

\texttt{>> ls -1 bin/*}

\texttt{field\_check}

\texttt{reform}

\texttt{rmt.x}
\\

If the doxygen package is installed \cite{doxygen}, the make command
produces a complete set of documents from the source code. These can be found in
\texttt{<path to repository>/rmt/docs/latex} (\LaTeX \, format) or \texttt{<path to
repository>/docs/html/index.html} (HTML format).

%%%%%%%%%%%%%%%%%%%%%%%%%%%%%%%%%%%%%%%%
\section{List of modules}
The main set of modules (under \texttt{/source/modules}) is compiled into the library \texttt{modules}. This library is then used for the compilation of both the main program, \texttt{rmt.x}, as well as various utility codes (see Sec. \ref{sec:executables}).

\begin{itemize}
\item
\texttt{angular\_momentum.f90}
Contains routines for calculating Clebsch-Gordan and Racah coefficients, as well as factorials.
\item
\texttt{calculation\_parameters.f90}
Contains parameters primarily for varying aspects of the
calculations, i.e., parameters which are not dependent on the system, the model of the system or the laser field.
\item
\texttt{checkpoint.f90}
Handles periodic write-outs of wavefunction data, calculation parameters and data arrays for calculation restart.
\item
\texttt{communications\_parameters.f90}
Defines all parameters required in parallel communication, specifically the numbers of processing elements to use in each region, the location of master cores and MPI communicators.
\item
\texttt{coordinate\_system.f90}
Routines for arbitrary solid rotation of the coordinate system. Used to change orientation of the polarization plane.
\item
\texttt{coupling\_rules.f90}
Handles all dipole (and other) coupling rules that are used in other modules, most notably \texttt{outer\_hamiltonian}.
\item
\texttt{distribute\_hd\_blocks.f90}
Handles the distribution of the inner-region atomic/molecular data (dipole blocks, state energies and boundary amplitudes) across processors. 
\item
\texttt{distribute\_hd\_blocks2.f90}
Handles the distribution of the inner-region atomic/molecular data
within each symmetry block. Each block has its own master which has already
received all the data from the inner-region master in
\texttt{distribute\_hd\_blocks}.
\item
\texttt{distribute\_wv\_data.f90}
Handles the distribution of the wavefunction data within each symmetry block.
\item
\texttt{eigenstates\_in\_kryspace.f90}
Handles the treatment of the Krylov subspace Hamiltonian for the Arnoldi propagation method.
\item
\texttt{electric\_field.f90}
Contains routines for calculating the electric field strength at a given instant of time.
\item
\texttt{fftpack.f90}
Routines from Netlib FFTPACK used in the calculation of photoelectron momentum distributions. All routines use \texttt{REAL(wp)} rather than the default \texttt{REAL} in order to be compatible with the RMT code.
\item
\texttt{file\_num.f90}
Handles the numerical portion of output/input file names.
\item
\texttt{finalise.f90}
Cleans up at the end of an RMT calculation. Frees memory, shuts down MPI communications, and closes output files.
\item
\texttt{global\_data.f90}
Standard UK-AMOR module handling frequently used constants.
\item
\texttt{global\_linear\_algebra.f90}
Handles linear algebra (inner products) on arrays, important tasks concerning the outer-region numerical grid and collates distributed population data.
\item
\texttt{grid\_parameters.f90}
Sets up outer-region grid and communications associated therewith.
\item
\texttt{initial\_conditions.f90}
Sets initial conditions for atomic/molecular system and laser parameters, parallel setup, checkpointing, etc.. Most parameters set in this module are read from \texttt{input.conf}.
\item
\texttt{initialise.f90}
High-level controller for setup of calculation. Calls routines to setup
communications, grid, timing, checkpointing, i-o, data arrays and to read in calculation data.
\item
\texttt{inner\_parallelisation.f90}
High-level controller for inner-region parallelisation.
\item
\texttt{inner\_to\_outer\_interface.f90}
Controls the flow of information from the inner to the outer region, specifically the matching of the wavefunction on the inner-region grid points. Note that a separate module, \texttt{outer\_to\_inner\_interface}, handles the flow of information in the other direction.
\item
\texttt{io\_files.f90}
Handles the opening and closing of output files, as well as the writing of files during the execution of the calculation. Data files written at the end of the calculation are handled in \texttt{io\_routines}.
\item
\texttt{io\_parameters.f90}
Sets parameters for the naming of output files.
\item
\texttt{io\_routines.f90}
General routines for writing data arrays to file. Used only at checkpoints of the calculation to output wavefunction and population data.
\item
\texttt{kernel.f90}
High-level controller for executing the Arnoldi propagation and setting up or sharing the data required for it.
\item
\texttt{krylov\_method.f90}
Handles all routines relevant to the Arnoldi propagator - setting up and sharing the reduced Hamiltonian, calling routines to step forward in time for both the inner and outer regions, etc.
\item
\texttt{kryspace\_taylor\_sums.f90}
Routines for checking the set up of the Krylov subspace.
\item
\texttt{live\_communications.f90}
Performs parallel operations relevant to the inner region. Sets up links between dipole-coupled symmetry blocks, and contains routines for performing the multiplication of the wavefunction by the Hamiltonian in parallel across the multilayer parallelized blocks.
\item
\texttt{local\_ham\_matrix.f90}
Routines for handling the outer-region wavefunction multiplication by the Hamiltonian. This includes incrementing with
the second derivative and Laplacian, applying the absorbing boundary conditions, and
calculating the ionized population. Incrementing with
the long-range potential matrices is handled in \texttt{outer\_hamiltonian} and
\texttt{outer\_hamiltonian\_atrlessthanb}.
\item
\texttt{lrpots.f90}
Sets up long-range potential matrices $W_E, W_P$ and $W_D$ for the outer region.
\item
\texttt{mpi\_layer\_lblocks.f90}
Sets up the first layer of parallelism in the inner region, entailing the division of the Hamiltonian matrix into symmetry blocks, and the allocation of processors to blocks based on their sizes. Also sets up the second layer of parallelism, establishing communicators within each block, and between the block masters.
\item
\texttt{outer\_hamiltonian.f90}
Manages operations involving the outer-region Hamiltonian (i.e., performing the Hamiltonian-wavefunction multiplication, incrementing with the long range
potential matrices).
\item
\texttt{outer\_hamiltonian\_atrlessthanb.f90}
Handles the outer-region Hamiltonian-wavefunction multiplication operations specifically on the inner-region grid.
\item
\texttt{outer\_to\_inner\_interface.f90}
Controls the flow of information from the outer region to the inner region, specifically the calculation of the boundary amplitudes for propagation of the outer-region wavefunction on the inner-region grid.
\item
\texttt{postprocessing.F90}
Calculates the full-dimensional photoelectron probability density and momentum distributions, as outlined in Sec.\;\ref{sec:reform}. 
\item
\texttt{potentials.f90}
Manages the provision of the long-range potential matrices during the calculation. The matrices are set up in \texttt{lrpots}, but accessed here.
\item
\texttt{precisn.f90}
Standard UK-AMOR module for setting parameters related to the numerical precision of the calculations.
\item
\texttt{propagators.f90}
Handles the outer-region Arnoldi propagation.
\item
\texttt{ql\_eigendecomposition.f90}
Module to find eigen-pairs for matrices by the QL method.
\item
\texttt{readhd.f90}
Reads the appropriate target structure and Hamiltonian files for the selected atomic/molecular mode,
and then sets up the large Hamiltonian for a time-dependent calculation.
\item
\texttt{rmt\_assert.f90}
Checks the truth of a statement and enables a graceful exit if false.
\item
\texttt{serial\_matrix\_algebra.f90}
Contains linear algebra routines for calculations involving the Krylov-subspace Hamiltonian.
\item
\texttt{setup\_bspline\_basis.f90}
Sets up the $B$-spline basis (in the atomic case) to be used in the inner region.
\item
\texttt{setup\_wv\_data.f90}
High-level controller for wavefunction data setup and tear-down (routines used only at the start and end of the calculation).
\item
\texttt{splines.f90}
Utility routines for the calculation of $B$-splines.
\item
\texttt{stages.f90}
Handles the position within a calculation (time-steps) when a calculation is starting or restarting.
\item
\texttt{tdse\_dependencies.f90}
Handles the dependencies of the main program (to keep the main program clean).
\item
\texttt{wall\_clock.f90}
Handles the timing of a calculation, including the writing out of the timing files.
\item
\texttt{wavefunction.f90}
High-level controller for the wavefunction files, including the reading and writing of wavefunction data, the extraction of observables (populations and expectation
values), and the application of an absorbing boundary.
\item
\texttt{work\_at\_intervals.f90}
Checks whether or not periodic tasks should be implemented on a given time step.
\end{itemize}

\subsection{A note on precision}\label{a-note-on-precision}

The level of precision used for floating point variables is 14 significant figures (double precision). The precision is set in the module {\texttt{precisn}} by adjusting the parameter {\texttt{decimal\_precision\_long}}, which is set to 15 by default. Care should be taken when adjusting this parameter, however, as the MPI communications assume double-precision variables are being used. 

\subsection{Executables}\label{sec:executables}
Besides the main executable, \texttt{rmt.x}, there are several utility codes which are compiled against the \texttt{modules} library. After compilation, these can be located in \texttt{<path\_to\_compile\_directory>/bin/}.

\begin{itemize}
    \item 
    \texttt{field\_check} Calculates the electric field defined by the
    parameters in a particular \texttt{input.conf} file. The output is a file,
    \texttt{EField.test}, identical to that produced by the full RMT calculation.
    \item
    \texttt{reform} Reconstruct the unformatted wavefunction files output by the RMT calculation (see Sec.\;\ref{sec:wavefunction-data}). Optional calculation of probability density and momentum distributions (see Sec.\;\ref{sec:reform}).
\end{itemize}

\section{Results}
\label{sec:results}
\subsection{Arbitrarily polarized light}

For many years, the use of time-delayed, counter-rotating, circularly polarized laser pulses has been of interest in a number of applications. The importance of such a configuration of pulses first became apparent in connection with high-harmonic generation and the synthesis of elliptically polarized, attosecond pulse trains \cite{Eichmannetal1995,long1995,Milosevicetal_coplanarmixing2000,MilosevicBecker_ellipticalaspulsetrains2000}. More recently, it has been found that photoelectron momentum distributions arising from such pulses carry novel signatures of wave-particle duality. Indeed, as predicted theoretically \cite{electron_vortices,rmt_arb}, and demonstrated experimentally \cite{pengel_prl,pengel_pra}, the distributions typically contain multi-armed spirals, caused by the interference between outgoing electron wavepackets with different magnetic quantum numbers. Here we investigate the response of a truly multielectron target, Ar, to counter-rotating circularly polarized pulses.

In the inner-region, the multielectron wavefunction is expanded in a basis of field-free \(R\)-matrix eigenfunctions. The Ar atom is described using the \(1s^2 2s^2 2p^6 3s^2 3p^5\) \(^2P^o\) and \(1s^2 2s^2 2p^6 3s 3p^6\) \(^2S^e\) residual-ion states of Ar\(^+\), to which we add a single electron that is subsequently ionized. The radial extent of the inner region is 20 a.u., and the inner-region continuum functions are generated using a set of 50 \(B\)-splines of order 9 for all angular momenta of the outgoing electron. We thus account for all \(1s^2 2s^2 2p^6 3s^2 3p^5 \epsilon l\) and \(1s^2 2s^2 2p^6 3s 3p^6 \epsilon l'\)  channels up to a maximum angular momentum \(L=L_{\rm max}\). Here we set \(L_{\max} = 12\), which is sufficient for a high degree of convergence at the laser parameters used in this case. The wavefunction is propagated for a total time of 500 a.u., which is around 15 laser cycles after the laser pulse terminates. This leaves ample time for ionizing wavepacket to proceed sufficiently far into the outer region to enable subsequent analysis. The maximum outer region radial extent is 1990 a.u., which is sufficient to contain the outgoing wavepacket during the propagation.

%%%%%%%%%%%%%%%%%%%%%%%%%%%%%%%%%%%%%
\begin{figure}[t]
    \centering
    \centerline{\includegraphics[trim={2cm 2cm 2cm 2cm},width=0.5\textwidth]{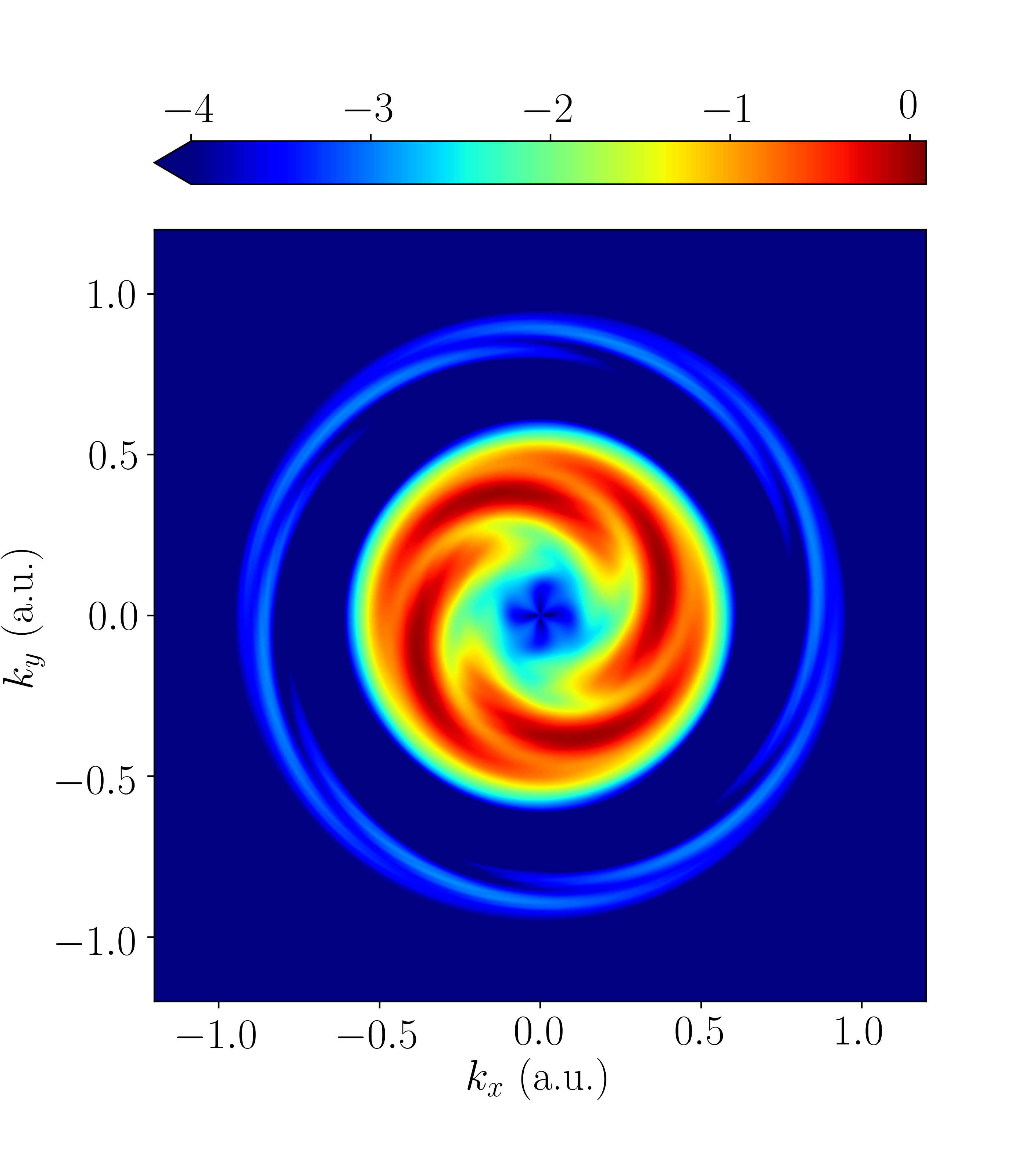}}
    \caption{Photoelectron momentum distribution for Ar, irradiated by a pair of counter-rotating, circularly polarized, 6-cycle, 9-eV, \(5\times 10^{13}\) Wcm$^{-2}$ laser pulses.}
    \label{ar8ev}
\end{figure}
%%%%%%%%%%%%%%%%%%%%%%%%%%%%%%%%%%%%%%%

%%%%%%%%%%%%%%%%%%%%%%%%%%%%%%%%%%%%%
\begin{figure}[ht]
    \centering
    \centerline{\includegraphics[width=0.8\textwidth]{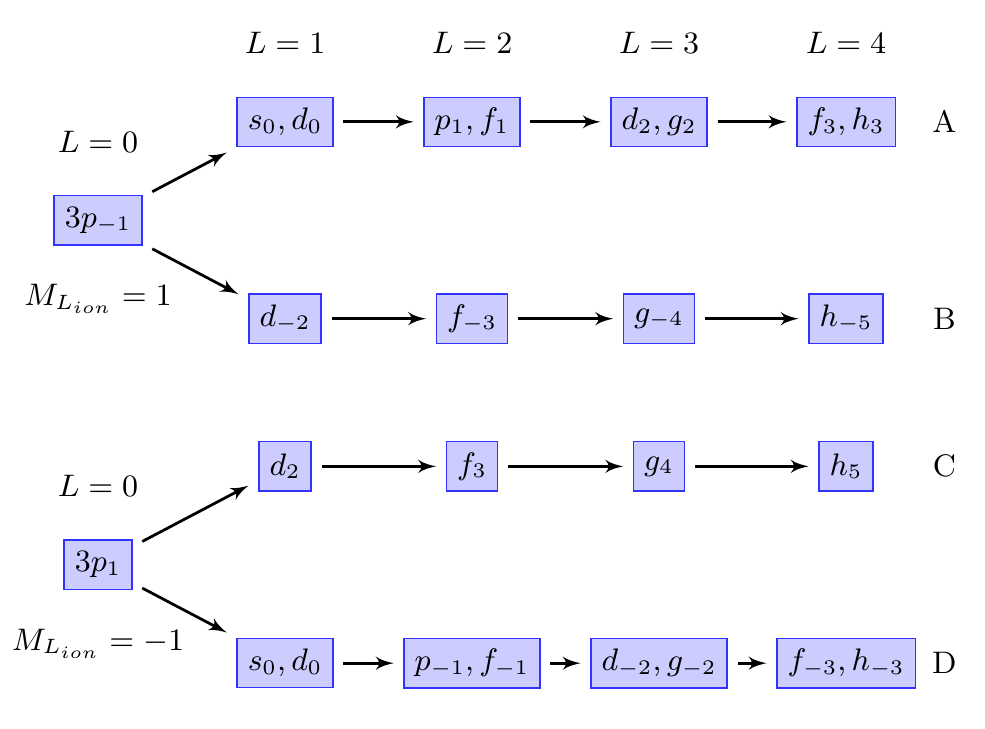}}
    \caption{Pathways for ionization of Ar \(3p\) electrons by a pair of counter-rotating, circularly polarized pulses. Electron \(l\) and \(m_l\) values are shown in boxes, and the total orbital angular momentum \(L\) and residual-ion magnetic quantum number \(M_{L_{\rm ion}}\) are also indicated.}
    \label{arpaths}
\end{figure}
%%%%%%%%%%%%%%%%%%%%%%%%%%%%%%%%%%%%%%%

Fig.\;\ref{ar8ev} shows the photoelectron momentum distribution, in the polarization plane ($k_x,k_y$), following irradiation of Ar by a pair of counter-rotating, circularly polarized, 9-eV ($\omega = 0.33$ a.u.), 6-cycle (3 cycles ramp-on, 3 cycles ramp-off), $5\times10^{13}$ Wcm$^{-2}$ laser pulses, with a relative delay of 2 fs. The distribution displays a four-arm spiral feature at low momenta ($k\approx0.4$ a.u.), arising from two-photon ionization.

Recent RMT calculations have observed similar features in the analogous distributions for two-photon ionization of He \cite{rmt_arb}. However, in the latter case, an initially bound $1s$ electron is ionized, whereas in Ar, either a \(3s\) or a \(3p\) electron is ejected. Furthermore, the ejected $3p$ electron can have magnetic quantum number $m_l=0,\pm1$. Since we consider laser pulses polarized in the $xy$ plane, contributions to the momentum distribution in this plane from $3p_0$ electrons may be eliminated by symmetry. Our calculations also demonstrate that ejection of a \(3s\) electron is strongly suppressed at the laser parameters considered in this case. Therefore, ejected $3p_{\pm1}$ electrons are the dominant constituents of the distribution shown in Fig.\;\ref{ar8ev}.

Since the initial symmetry of neutral Ar is \(S^e\), \(3p_{\pm1}\) electrons must couple to a residual ion with magnetic quantum number \(M_{L_{\rm ion}}=\mp1\). With this in mind, four ionization pathways emerge, two for each electron-ion coupling, as shown in Fig.\;\ref{arpaths}. In paths A and B, a \(3p_{-1}\) electron absorbs photons from each pulse, with their opposite helicities progressively increasing \(m_l\) on path A, and decreasing \(m_l\) on path B. After two photon absorptions, a superposition of \(p_1,f_1\), and \(f_{-3}\) continuum-electron wavepackets is created, whose interference yields a four-arm spiral pattern in the momentum distribution for these pathways. Similarly, in paths C and D, two photon absorptions yield continuum electrons with \(m_l=-1\) and \(m_l=3\), which interfere to give another four-arm spiral. Since the residual-ion contributions are summed incoherently (see Eq.\;\eqref{momdiseq}, Sec.\;\ref{sec:reform}), the total ($k_x,k_y$) distribution will take the form of a four-arm spiral at $k\approx0.4$ a.u., corresponding to the momentum attained following absorption of two 9-eV photons.

At this intensity, three-photon (above-threshold) ionization is also significant, resulting in an outer, six-arm spiral feature (at around $k\approx0.9$ a.u.). This arises due to the interference of $g_2$ and $g_{-4}$ electrons on paths A and B, and $g_4$ and $g_{-2}$ electrons on paths C and D.

\subsection{Relativistic atomic systems}

\begin{figure}[ht]
    \centering
    \centerline{\includegraphics[height=6cm]{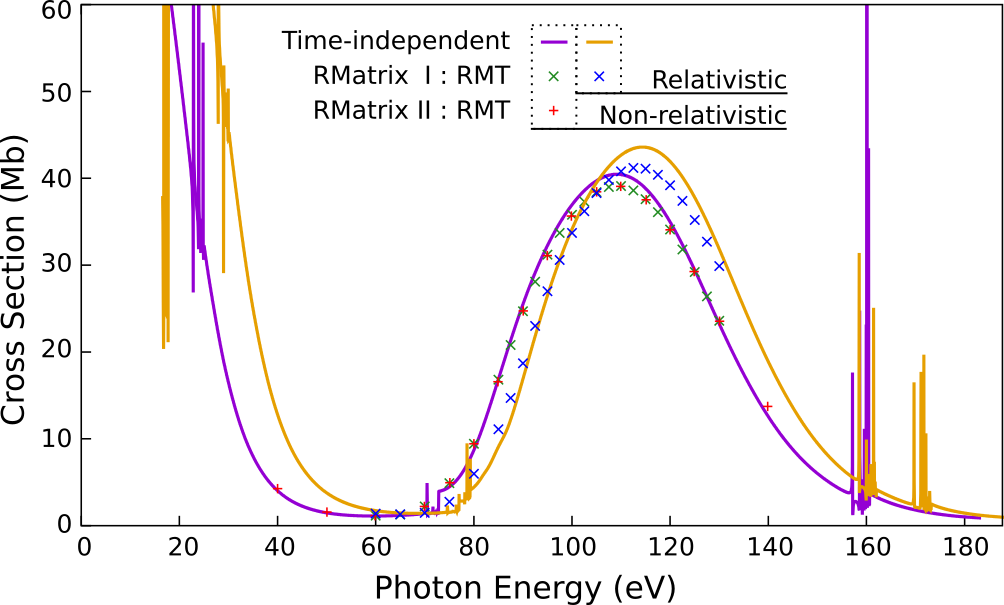}}
    \caption{Single-photon ionization cross-sections for atomic xenon, as calculated using time-independent (solid lines) and time-dependent (individual points) R-matrix approaches. The RMT calculations employ a 12-cycle pulse (3 cycles ramp-on, 6 cycles at maximum intensity and 3 cycles ramp-off) of peak intensity $0.13 \times 10^{14}$ Wcm$^{-2}$. The relativistic calculations (yellow line and blue `$\times$' symbols) are those including the spin-orbit interaction. The non-relativistic data (green `$\times$' symbols and purple line) are obtained using a $jK$ coupling scheme, but with the spin-orbit interaction excluded. The non-relativistic RMatrix II results (red `+' symbols) are computed in $LS$ coupling.
    \label{fig:XenonSinglePhoton}}
\end{figure}

%JW version:

%Figure \ref{fig:XenonSinglePhoton} shows example calculations of the single photon ionization cross section of Xenon. The results are obtained using a model of atomic xenon constructed from residual ions that describe ejection from the 4p, 4d, 5s and 5p shells. We present both non-relativistic results (to enable comparison between the RMatrixII and RMatrixI methods) and semi-relativistic data with the spin-orbit interaction included. Both sets of atomic data are described using $jK$ coupling.

%All time-dependent RMT cross-section data points are obtained using a 12 cycle pulse (3 cycles ramp-on, 6 cycles at maximum intensity and 3 cycles ramp-off) of intensity $0.13 \times 10^{14}$ Wcm$^{-2}$. There is strong agreement between the non-relativistic RMatrixI and RMatrixII cross section values, as would be expected. For further comparison we present a cross-section spectrum calculated from the same RMatrixI xenon model using the RMatrixI time-independent outer region routines. Here again there is strong agreement, after the effect of the short pulse is taken into account. As the pulse length increases, the time-dependent results will converge to the time-independent cross section. 

%To demonstrate the effect of the spin-orbit interaction, we run RMT and the RMatrixI outer region codes using a semi-relativistic atomic xenon model to get  time-dependent and time-independent single-photon ionization cross-section spectra. Again there is good agreement.

%DDAC version:

To demonstrate the capabilities of our semi-relativistic RMT methodology, we perform calculations for the photoionization cross-section of atomic Xe in the XUV spectral range. The Xe system is a natural candidate for study, not only by virtue of its high mass ($Z = 54$), but also the enhanced relative importance of multielectron correlations in its photoionization dynamics, as manifested by the so-called giant dipole resonance \cite{amusia_2000,multielectron_atoms}.

The Xe target considered here is as described in a previous RMT study \cite{ola_nearIR}. In the inner region, we regard the atom as Xe$^+$ to which is added a single electron. To describe the structure of Xe$^+$, we employ a set of Hartree-Fock orbitals, acquired for the ionic ground state from the data of Ref. \cite{clementi_atomic_data}. These orbitals are used to construct the $4p^{-1}$, $4d^{-1}$, $5s^{-1}$ and $5p^{-1}$ states of the Xe$^{+}$ ion.

The radial extent of the inner region is 20 a.u., which suffices to effectively confine the orbitals of the Xe system. The inner-region continuum functions are generated using a set of 50 $B$-splines of order 13 for each available orbital angular momentum of the outgoing electron. We retain all admissible ionization channels up to a maximum total orbital angular momentum of $L_{\rm max} = 5$ when relativistic corrections are excluded, and  a maximum total angular momentum of $J_{\rm max} = 5$ when they are retained. The multielectron wavefunction is propagated for a total time of 2000 a.u., ensuring that the populations of the single-photon ionization channels have stabilized, and thus that the photoionization cross-section is reliably determined. The outer-region boundary radius is approximately 3500 a.u., such that no unphysical wavepacket reflections occur during the wavefunction propagation.

Fig. \ref{fig:XenonSinglePhoton} presents the calculated single-photon ionization cross-section of Xe. There, we display both non-relativistic cross-section data (to enable a comparison between calculations employing RMatrixI and RMatrixII atomic-structure input), as well as semi-relativistic cross-section data incorporating the spin-orbit correction. Note that, for RMatrixI input data, we also draw comparison between the results of a time-dependent RMT simulation, and time-independent calculations following the standard $R$-matrix theory of photoionization, as implemented in the RMatrixI package \cite{belfastatomic}. For our time-dependent RMT simulation, we have assumed a 12-cycle pulse (3 cycles ramp-on, 6 cycles at maximum intensity and 3 cycles ramp-off) of peak intensity $1.3 \times 10^{13}$ Wcm$^{-2}$.

All calculations suggest that the photoionization spectrum of Xe is dominated by a broad peak, which we identify as the giant dipole resonance. The latter arises from the excitation of a $4d$ electron into a transient, quasi-bound state of $f$ character~\cite{Beckeretal1989_Xe}. However, this feature cannot be understood purely as a single-particle phenomenon: partial photoionization spectra evidence the effect of the resonance in photoemission from the $5s$ and $5p$ subshells, as well as from the $4d$ subshell. It should be noted that the observed position of the resonance in each case -- around 110 eV -- differs markedly from that found in experimental photoionization spectra, therein appearing around 100 eV~\cite{Beckeretal1989_Xe}. This discrepancy can be traced to the quality of the Xe target models. In general, in order to produce accurate RMT data, the R-matrix basis set should retain the most important configuration-interaction effects. In particular, for the photoionization of Xe, the $4d^{2} \rightarrow 4f^{2}$ double excitations are especially relevant in describing the Xe$^+$ residue in the $4d^{9}5s^{2}5p^{6}\epsilon f$ channel. For the purpose of demonstrating our semi-relativistic RMT method, we have simply omitted these excitations in our analysis, preferring to use readily available data for the Hartree-Fock orbitals of the Xe$^+$ ground state \cite{clementi_atomic_data}.\ Additional, suitably optimized orbitals for more sophisticated simulations could be obtained, for example, from multiconfiguration Hartree-Fock calculations \cite{FroeseFischer_MCHF}. More crucially, however, the configurations included for the description of the $N$-electron system, using the traditional RMatrixI and RMatrixII computer codes, suffer a certain limitation: when generating $(N+1)$-electron configurations, the occupancy of $f$-subshells must be restricted to at most two electrons. As such, $4d^{2} \rightarrow 4f^{2}$ double promotions cannot be retained in a configuration-interaction expansion for the $4d^{-1}$ state of Xe$^+$, since these would require the corresponding $4d^{7}5s^{2}5p^{6}4f^{3}$ configuration for completeness. The poorer representation of the Xe$^+$ $4d^{-1}$ residual ion gives rise to an artificially higher ionization threshold relative to the Xe ground state, and ultimately, a less accurate total wavefunction. This poorer accuracy is reflected in Fig \ref{fig:XenonSinglePhoton}, where for all data sets, the giant resonance is shifted to higher energies by about 10 eV. A similar effect has also been reported by Gorczyca {\it et al.} \cite{Gorczyca_endohedralXe}.

Excellent agreement is found between calculations employing the non-relativistic, RMatrixI and RMatrixII atomic-structure data, as should be the case. Moreover, we find a high degree of qualitative concurrence between the time-dependent and time-independent computations employing RMatrixI input. The quantitative discrepancies are largely attributable to the finite pulse length adopted in the RMT calculations, which contrasts the ideally monochromatic field assumed in the time-independent analysis. Indeed, the level of agreement has been found to improve systematically with increasing pulse length. Finally, our results appear to suggest that the Breit-Pauli corrections have only a minor effect on the photoionization spectrum in this energy range.

\subsection{Molecules}
%DDAC: Modified discussion below.
To illustrate the use of the molecular RMT approach, we have performed calculations for  two- and four-photon ionization of H$_2$ at the  equilibrium internuclear distance, 1.4~a.u.. The aim was to reproduce previous R-matrix-Floquet calculations \cite{RMF_H2}.  Here, the laser radiation is linearly polarized along the \(z\)-axis (\(\bm{E}(t) = (0, 0, E_z(t))\)), which also coincides with the molecular axis. In the photon energy range investigated, only two states of H$_2^+$, namely the ground state $^2\Sigma_g^+$ ($^2$A$_g$ in the D$_{2h}$ point group), and the lowest-lying excited state $^2\Sigma_u^+$ ($^2$B$_{1u}$), need to be considered.

%To illustrate the capabilities of the molecular RMT approach, a subset of the stationary Floquet R-matrix results for two- and four-photon  photoionization of H$_2$ at equilibrium internuclear distance 1.4~a.u. \cite{RMF_H2} was recalculated with RMT and is shown in Figs.~\ref{fig:H2-2photon} and \ref{fig:H2-4photon}. Here, the laser radiation is linearly polarized in the \(z\)-axis (\(\bm{E}(t) = (0, 0, E_z(t))\)), which also coincides with the molecular axis. In the photon energy range investigated, only two states of H$_2^+$, the ground state $^2\Sigma_g^+$ ($^2$A$_g$ in the D$_{2h}$ point group the calculations are performed in) and the lowest lying $^2\Sigma_u^+$ ($^2$B$_{1u}$) need to be considered.  

The molecular orbitals were obtained for H$_2^+$  using a hand-crafted Gaussian basis centered on each of the hydrogen atoms (see Table~\ref{tab:H2-basis}). These were used in the UKRmol+ suite to generate the inner region data. The configurations involved in describing the electronic states of both H$_2$ and H$_2^+$ are summarized in Table~\ref{tab:H2-orbs}. 
The continuum was modelled  using a center-of-mass-centered $B$-spline  basis \cite{molecular_continuum} extending up to \(R_a = 30\)~a.u., the radius of the R-matrix sphere, and consisting of 45 $B$-splines of order 6. The symmetric orthogonalization employed a deletion threshold of 10$^{-5}$.

\begin{table}[ht]
    \centering
    \begin{tabular}{ll}
        \toprule
        Orbital & Gaussian exponents \\
        \midrule
        \(1s\) & 23.10, 4.240, 1.190, 0.407, 0.1580 \\
        \(2s\) & 55.17, 4.640, 1.428, 0.190, 0.0711 \\
        \(2p\) & 52.91, 9.655, 2.796, 1.006, 0.4013, 0.1677 \\
        \bottomrule
    \end{tabular}
    \caption{Gaussian exponents, optimized to model the three-function basis of Slater-type orbitals (STOs) employed in Ref. \cite{RMF_H2}, for the sample H$_2$ calculation. The original STO basis could not be used, since UKRmol+ supports only Gaussian-type orbital bases for bound orbitals.}
    \label{tab:H2-basis}
\end{table}

\begin{table}[ht]
    \centering
    \begin{tabular}{llll}
        \toprule
        Molecule & State      & Energy (a.u.) & Electronic configurations \\
        \midrule
        H$_2^+$  & \(1\,{}^2\Sigma_g^+\)  (\(1 {}^2\!A_g\))    & \(-0.56777\) & \(1a_g^1\)    \\
%                 & \(1B_{1u}\) & \(+0.11818\) & \(1b_{1u}^1\) \\
        \midrule
        H$_2$    & \(1\,{}^1\Sigma_g^+\)  (\(1\,{}^1\!A_g\))    & \(-1.15279\) & \(1a_g^2\), \(1b_{1u}^2\), \(1a_g^1 \epsilon a_g^1\), \(1b_{1u}^1 \epsilon b_{1u}^1\) \\
                 & \(1\,{}^1\Sigma_u^+\)  (\(1\,{}^1\!B_{1u}\)) & \(-0.69964\) &  \(1a_g^1 1b_{1u}^1\), \(1a_g^1 \epsilon b_{1u}^1\), \(1b_{1u}^1 \epsilon a_g^1\) \\
                 & \(2\,{}^1\Sigma_u^+\)  (\(2\,{}^1\!B_{1u}\)) & \( -0.62531\) &  \textit{ditto} \\
                 & \(3\,{}^1\Sigma_u^+\)  (\(3\,{}^1\!B_{1u}\)) & \(-0.60217 \) & \textit{ditto} \\
        \bottomrule
    \end{tabular}
    \caption{Configurations used to build the electronic wavefunctions in the sample H$_2$ calculation; $\epsilon$ denotes an orbital built mainly from continuum functions. The calculated ground-state energies agree well with the values \(-0.569246\)~a.u. and \(-1.152682\)~a.u. reported in Ref. \cite{RMF_H2}. The labels of the states are provided for both the actual point group of H$_2$ (D$_{\infty h}$) and the one used in the calculations (D$_{2h}$).}
    \label{tab:H2-orbs}
\end{table}

\begin{figure}
    \centering
    \centerline{\includegraphics{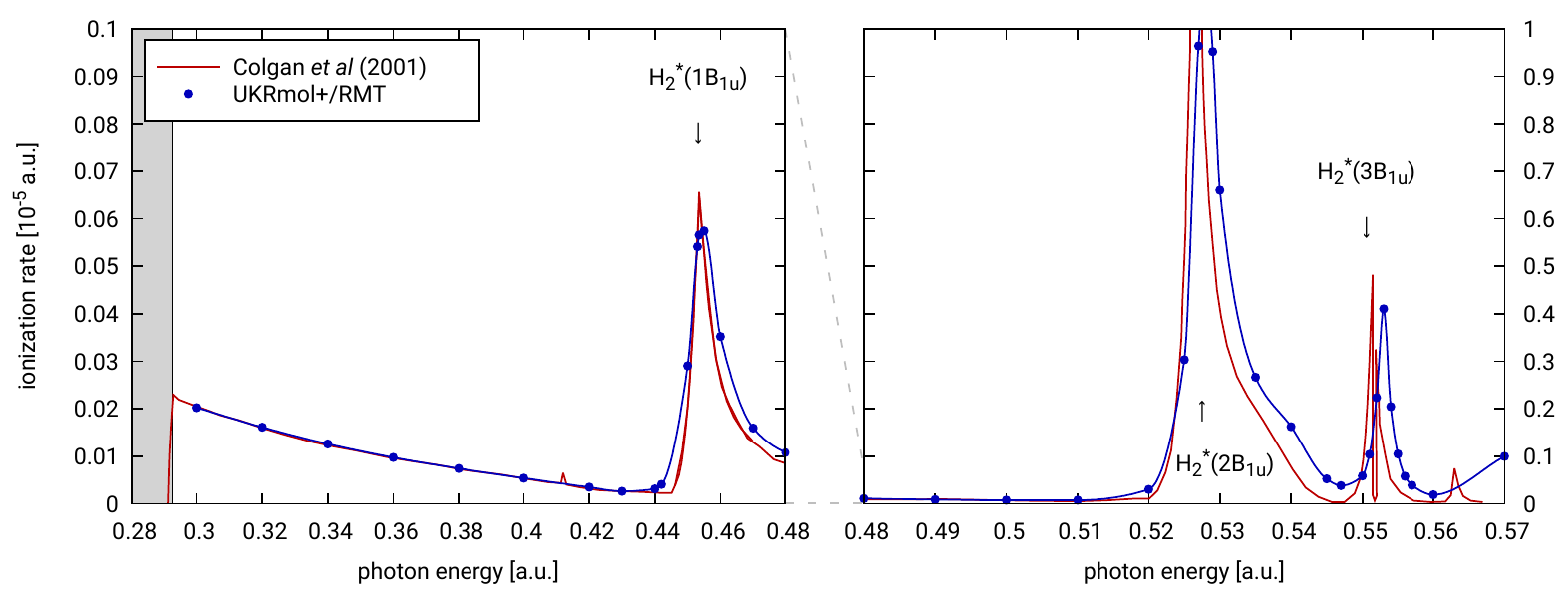}}
    \caption{Stationary ionization rate for H$_2$ with fixed internuclear distance 1.4~a.u., subject to a linearly polarized laser pulse with intensity \(I = 10^{12}\)~W/cm$^2$, between the two-photon and one-photon ionization thresholds (white area). Current results are compared to those of previous R-matrix-Floquet calculations reported by Colgan {\it et al.} \cite{RMF_H2}.} %The indicated H$_2$ B$_{1u}$ states of D$_{2h}$ correspond to $\Sigma_u^+$ in the D$_{\infty h}$ point group.
    \label{fig:H2-2photon}
\end{figure}

\begin{figure}
    \centering
    \centerline{\includegraphics{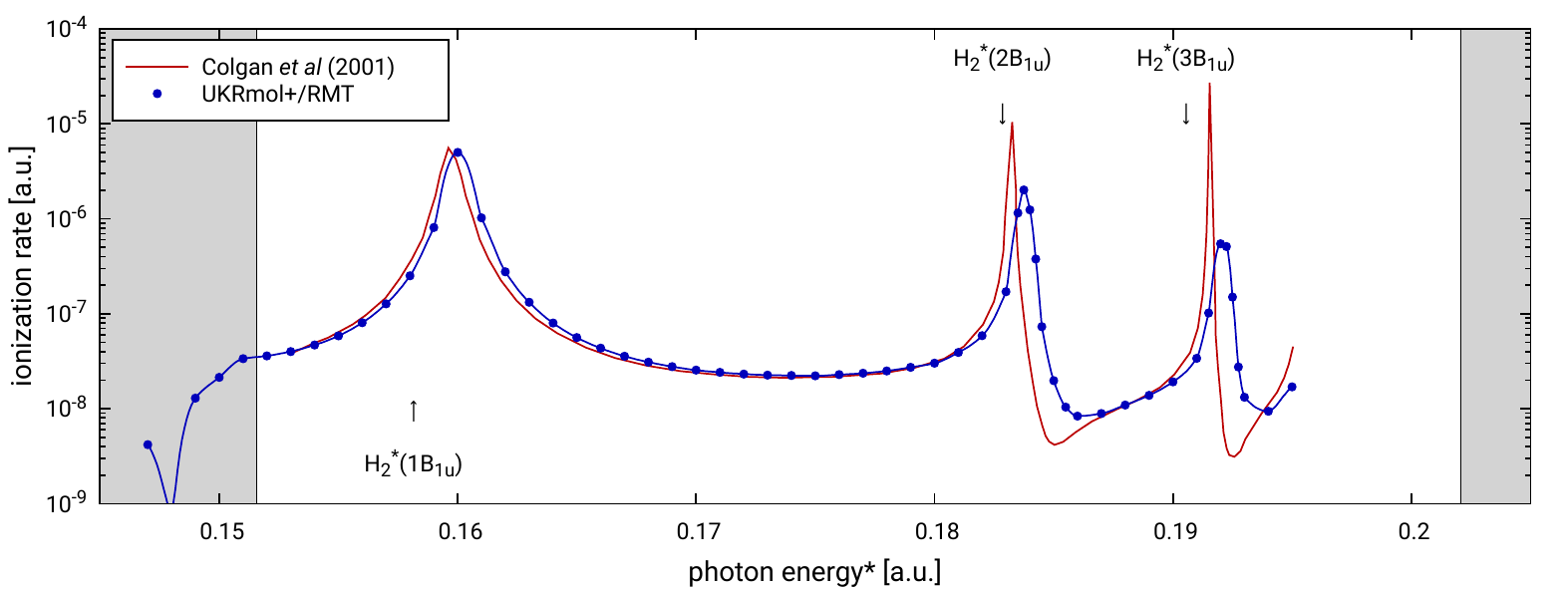}}
    \caption{As in Fig.~\ref{fig:H2-2photon}, but with \(I = 10^{13}\)~W/cm$^2$, and now between the four-photon and three-photon ionization thresholds (white area). For the purpose of this particular figure, the ground-state energy of H$_2$ was shifted to agree with the value \(-1.173949\)~a.u. given in Ref.~\cite{Sharp1971}.}
    \label{fig:H2-4photon}
\end{figure}

The total time-dependent population in the outer region beyond a selected distance (80 a.u.) was monitored for a sufficiently long time, typically  thousands of atomic units, but always short enough to allow just a small fraction of the electronic charge to leave the inner region. The total ionization rate was then obtained as an average increase of the outer population per unit of time (linear fit), and is plotted in Figs.~\ref{fig:H2-2photon} and \ref{fig:H2-4photon}. Away from the resonances, the time-dependent (RMT) results agree remarkably well with those of the time-independent (UKRmol+) calculations. Near the resonances, however, and particularly those close to the one-photon ionization threshold at \(E_{IP} = 0.585\)~a.u., there are noticeable deviations between the RMT and the earlier R-Matrix-Floquet results. This is likely a consequence of the shorter range of our Gaussian-type orbitals, as compared to the slowly decaying Slater-type orbitals used in Ref. \cite{RMF_H2}. Nevertheless, the positions of the resonances agree sufficiently well to confirm that the excitation thresholds of H$_2$ (not explicitly listed in Ref. \cite{RMF_H2}) are reproduced with fair accuracy.

We also present the four-photon results obtained by the same method, at higher intensity \(I = 10^{13}\)~W/cm$^2$, in Fig.~\ref{fig:H2-4photon}. Again, good agreement with the R-matrix-Floquet calculations is observed for low energies, while higher resonances deviate somewhat from those in Ref. \cite{RMF_H2}. We ascribe this, in the first instance, to the differences in the modelling of the molecular states, but additional differences may result from the details of the implementation on the R-matrix-Floquet and RMT methods, such as the use of the velocity gauge in the outer region in the Floquet code.

%Here the displacement of resonances is non-negligible. Assuming that the excitation thresholds in the R-matrix Floquet calculation \cite{RMF_H2} were the same as in the present model, which is supported by the agreement in positions of the two-photon resonances, the disagreement in case of four-photon ones must be caused by a different description of the interaction of the molecule with the field. A weakly bound excited state receives, to a good approximation, positive energy contribution from the mean kinetic energy transferred from the field to the electron per a cycle [Springer handbook, \S 74.2],
%\begin{equation}
%    \Delta E = 2\pi\alpha (1 + \varepsilon) \frac{I}{\omega^2} \,,
%    \label{eq:ponderomotive-shift}
%\end{equation}
%where \(\alpha\) is the fine structure constant, \(\varepsilon\) the ellipticity (zero in this calculation), \(I\) is the irradiance and \(\omega\) the photon energy of the laser field; atomic units are assumed for all quantities. For lower excited states, this formula represents an upper bound on the energy shift, because the field does not affect these states as much. As indicated in Fig.~\ref{fig:H2-4photon}, the shift does not allow for such a big offset as reported in Ref. \cite{RMF_H2}, raising doubts over the inner consistency of their results, but supporting the present calculations.

%Green arrows indicate maximal field-induced shifts of the excitation thresholds of H$_2$ (resonance positions) due to Eq.~\eqref{eq:ponderomotive-shift}.

\section{Acknowledgements}
The RMT code is part of the UK-AMOR suite, and can be obtained for free from Ref. \cite{RMT_repo}. The results shown used the ARCHER UK National Supercomputing Service (www.archer.ac.uk), for which access was obtained via the UK-AMOR consortium funded by EPSRC. The development of RMT has benefited from computational support from CoSeC, the Computational Science Centre for Research Communities, through CCPQ. The authors acknowledge funding from the EPSRC under Grants No. EP/P022146/1, No. EP/P013953/1, and No. EP/R029342/1. ZM acknowledges partial support by OP RDE project No.CZ.02.2.69/0.0/0.0/16\_027/0008495, International Mobility of Researchers at Charles University.

The authors would also like to acknowledge the contribution of Michael Lysaght,
Robert McGibbon, Laura Moore, Lampros Nikolopoulos, Jonathan Parker, Martin Plummer and Ken Taylor to the RMT code.

\appendix
\newcommand{\chnl}[2]{\overline{\Phi}_{#1}^{\Gamma_{#2}}({\mathbf{X}}_{N};{\hat{\mathbf{r}}}_{N+1}\sigma_{N+1})}
\newcommand{\chnlshort}[2]{\overline{\Phi}_{#1}^{\Gamma_{#2}}}
\newcommand{\rsh}[1]{{X}_{l_{#1},m_{#1}}}
\newcommand{\tgt}[1]{\Phi_{#1}({\mathbf{X}}_{N})}

\section{Molecular outer-region potentials}\label{sec:appA}
The potentials to be derived here describe the interaction in the outer region of the ionized electron with the combined field of the laser and the residual, $N$-electron molecular target. The Schr\"{o}dinger equation describing the laser-molecule interaction is
\begin{eqnarray}\label{eq1}
\imath\frac{\partial}{\partial t}\Psi(\mathbf{X}_{N+1},t) = [H_{N+1} + D_{N+1}(t)] \Psi(\mathbf{X}_{N+1},t),
\end{eqnarray}
where $\mathbf{X}_{N+1}=\mathbf{x}_{1},\mathbf{x}_{2},\dots,\mathbf{x}_{N+1}$ with $\mathbf{x}_{i}=\mathbf{r}_{i}\sigma_{i}$ stands for all space and spin coordinates of the $N+1$ electrons of the molecule, $H_{N+1}$ is the field-free non-relativistic Hamiltonian of the molecule in the fixed-nuclei approximation and $D_{N+1}(t)$ describes the interaction of the electrons with the laser field. In order to derive the outer-region potentials, it is convenient to rewrite $H_{N+1}$ and $D_{N+1}(t)$ in a form where the terms pertaining the $(N+1)$th electron are separated out:
\begin{eqnarray}
H_{N+1} &=& H_{N} + H_{1},\\
H_{N} &=& \sum_{i=1}^{N}\left( -\frac{1}{2}\nabla_{i}^{2} + \sum_{i>j}^{N}\frac{1}{\vert \mathbf{r}_{i} - \mathbf{r}_{j}\vert} -\sum_{k=1}^{Nuclei}\frac{Z_{k}}{\vert \mathbf{r}_{i} - \mathbf{R}_{k}\vert}\right),\\ 
H_{1} &=& -\frac{1}{2}\nabla_{N+1}^{2} + \sum_{i=1}^{N}\frac{1}{\vert \mathbf{r}_{N+1} - \mathbf{r}_{i}\vert}-\sum_{k=1}^{Nuclei}\frac{Z_{k}}{\vert\mathbf{r}_{N+1} - \mathbf{R}_{k}\vert}.\label{we}
\end{eqnarray}
Similarly, for the laser-electron interaction term in the dipole approximation and the length gauge, we have:
\begin{eqnarray}
D_{N+1}(t) &=& D_{N}(t) + D_{1}(t),\\
D_{N}(t) &=& \mathbf{E}(t)\cdot\sum_{i=1}^{N} \mathbf{r}_{i},\\ 
D_{1}(t) &=& \mathbf{E}(t)\cdot\mathbf{r}_{N+1}\label{efield},
\end{eqnarray}
where $\mathbf{E}(t)$ is the electric field as defined in Eq.~(\ref{def:efield}). In deriving the equations for the potentials we follow the notation of Refs.\;\cite{rmt_arb,burke}.

\subsection{Definition of the outer-region potentials}
The one-electron, radial potentials in the outer region are expressed in the basis of the channel functions $\chnlshort{p}{p}$, which are defined as:
\begin{eqnarray}\label{def:chnl}
\chnl{p}{p} &=& \sum_{M_{S_{p}}}\sum_{\mu_{p}} (S_{p}M_{S_{p}},\frac{1}{2}\mu_{p}\vert SM)\Phi_{p}({\mathbf{X}}_{N})\nonumber\\
&\times&\rsh{p}(\hat{\mathbf{r}}_{N+1})\chi_{\frac{1}{2},\mu_{p}}(\sigma_{N+1}),
\end{eqnarray}
where $\Gamma_{p}$ stands for the total spatial symmetry (irreducible representation) of the $(N+1)$-electron molecular wavefunction, $(S_{p}M_{S_{p}},\frac{1}{2}m_{p}\vert SM)$ is the Clebsch-Gordan coefficient ensuring correct spin-coupling of the target state $\Phi_{p}({\mathbf{X}}_{N})$ and the unbound electron to the total spin $S$. However, spin plays only a passive role in the molecular calculations, since relativistic corrections are not implemented in  UKRmol+  so we can assume that only wavefunctions corresponding to the same total spin have been included in the calculation. Therefore, we have omitted the spin variables on the left-hand side of the equation. The channel wavefunctions $\chnlshort{p}{p}$ are labelled by the collective index $p\equiv \{p_{t},l_{p},m_{p}\}$, where $\rsh{p}(\hat{\mathbf{r}}_{N+1})$ is a \textit{real} spherical harmonic~\cite{homeier1996} with angular-momentum quantum numbers chosen such that when coupled to the target state $\Phi_{p_{t}}({\mathbf{X}}_{N})$, the total spatial symmetry of the channel wavefunction is $\Gamma_{p}$.

%We note that as opposed to the atomic case the wavefunction of total symmetry $\Gamma_{p}$ and the target states $\Phi_{p_{t}}({\mathbf{X}}_{N})$ do not posses a definite value of angular momentum and its projection on the z-axis: in case of linear molecules the projection of the angular momentum on the molecular axis is a conserved quantity but the molecular UKRmol+ codes only use Abelian point groups which do not include point groups of linear molecules. Instead, the total molecular wavefunction is written as a sum of up to 8 wavefunctions each transforming according to an irreducible representation of the corresponding Abelian point group of the molecule. Concretely, in the outer region we have
Using the channel wavefunctions defined above, the total wavefunction with one electron in the outer region is written as
\begin{eqnarray}\label{eq:psi}
\Psi({\mathbf{x}}_{N+1},t) = \sum_{p}\chnl{p}{p}\frac{1}{r}f_{p}(r,t),
\end{eqnarray} 
where $f_{p}(r,t)$ is the time-dependent reduced radial wavefunction of the outer-region electron in channel $p$, and $r\equiv r_{N+1}$. Inserting~(\ref{eq:psi}) into Eq.~(\ref{eq1}) and projecting onto the channel wavefunctions, we obtain:
\begin{eqnarray}
&&\imath \frac{\partial}{\partial t}f_{p}(r,t) = h_{p}(r)f_{p}(r,t) + \sum_{p'=1}^{n}V_{pp'}(r,t)f_{p'}(r,t),\\
&&h_{p}(r) = -\frac{1}{2}\frac{d^2}{dr^2} + \frac{l_{p}(l_{p}+1)}{2r^2} + E_{p},\label{eq:hp}
\end{eqnarray}
where $E_{p}$ and $l_{p}$ are the energy of the residual, $N$-electron molecular state, and the continuum angular momentum corresponding to channel $p$, respectively. Also, $n$ is the total number of outer-region channels. The matrix $V_{pp'}(r,t)$ of the outer region potentials has three contributions
\begin{eqnarray}\label{eq:Vpot}
\mathbf{V}(r,t) = \mathbf{W}^{E}(r) + \mathbf{W}^{D}(t) + \mathbf{W}^{P}(r,t).
\end{eqnarray}
In the following, we will derive expressions for each of them. Since there are no spin-dependent interactions in our molecular calculations, and we include only wavefunctions corresponding to the same total spin $S$, the integration and summation over the spin-dependent variables is readily performed in the expressions for the potential matrix elements, with the trivial result~$\delta_{S,S}=1$. Therefore the spin-dependent part of the channel functions does not appear in the subsequent derivations.

\subsection{$W^{E}$ potentials for electron-target interaction}\label{sec:appendix1}
The potentials $W^{E}$ describe the field-free interaction of the electron with the residual molecular target (typically a positive ion), and are generated by UKRmol+ for electron-molecule scattering calculations. These potentials are matrix elements of the Coulombic terms from the Hamiltonian (\ref{we}), and their final form is:
\begin{equation}
W_{pp'}^{E}(r) = \sum_{\lambda = 0}^{\infty}a_{pp'\lambda}r^{-\lambda-1}, \mbox{ $p,p'=1,\dots,n$, $r \geq a$,}\label{ap1:eq:couplings}
\end{equation}
where $a$ is the R-matrix radius. The $W_{pp'}^{E}$ potentials are obtained by summing the electronic and the nuclear contributions:
\begin{eqnarray}
W_{pp'}^{E}(r) &=& W_{el,pp'}^{E}(r) + W_{nuc,pp'}^{E}(r),\\
W_{el,pp'}^{E}(r) &=& \bigg(\chnlshort{p}{}\bigg\vert \sum_{i=1}^{N}\frac{1}{\vert \mathbf{r}_{N+1} - \mathbf{r}_{i}\vert} \bigg\vert \chnlshort{p'}{}\bigg)\label{eq:We:el},\\
W_{nuc,pp'}^{E}(r) &=& \bigg(\chnlshort{p}{}\bigg\vert -\sum_{k=1}^{Nuclei}\frac{Z_{k}}{\vert\mathbf{r}_{N+1} - \mathbf{R}_{k}\vert}\bigg\vert \chnlshort{p'}{}\bigg)\label{eq:We:nuc}.
\end{eqnarray}
The round brackets indicate integration over all spin-space coordinates with exception of the radial coordinate of the $(N+1)$th electron. Note that the $W^{E}$ potentials are diagonal in the irreducible-representation index $\Gamma$, since the channel wavefunctions of different total symmetries are not coupled by the field-free Hamiltonian.

The derivation of $W_{pp'}^{E}(r)$ is based on the expansion of the Coulomb potential in Legendre polynomials $P_{\lambda}(\cos\theta)$: %\cite{friedrich2006}:
\begin{eqnarray}
\frac{1}{\vert \mathbf{r}_{i}-\mathbf{r}_{j}\vert}&=&\sum_{\lambda=0}^{\infty} \frac{{r_j}^{\lambda}}{r_{i}^{\lambda+1}}P_{\lambda}(\hat{\mathbf{r}}_{i}.\hat{\mathbf{r}}_{j})=
\nonumber\\
&=&\sum_{\lambda=0}^{\infty}\frac{4\pi}{2\lambda+1} \frac{{r_j}^{\lambda}}{r_{i}^{\lambda+1}}\sum_{m=-\lambda}^{\lambda}X_{\lambda,m}(\hat{\mathbf{r}}_{i})X_{\lambda,m}(\hat{\mathbf{r}}_{j}),\mbox{ $r_{j}\leq r_{i}$,} \label{ap1:eq:leg}
\end{eqnarray}
where $X_{\lambda,m}(\hat{\mathbf{r}})$ are real spherical harmonics. In the second step, we made use of the addition theorem for real spherical harmonics~\cite{homeier1996}, which decouples the angular variables of the two electrons. In the outer region, the distance of the unbound electron from the origin is always greater than that of the target molecule's electrons ($r_{N+1}>r_{i}$, $i=1,\dots,N$) and nuclei. Therefore, in the expressions below, the radial distance $r$ of the $(N+1)$th electron will always be in the denominator.

Using the Legendre expansion (\ref{ap1:eq:leg}), we calculate first the contribution of the electronic term (\ref{eq:We:el}):
\begin{eqnarray}
W_{el,pp'}^{E}(r)&=&\sum_{\lambda=0}^{\infty}\frac{4\pi}{2\lambda+1}\frac{1}{r^{\lambda+1}}\sum_{i=1}^{N}\left(\chnlshort{p}{}\bigg\vert\sum_{m=-\lambda}^{\lambda}X_{\lambda,m}(\hat{\mathbf{r}}_{N+1})X_{\lambda,m}(\hat{\mathbf{r}}_{i})r_{i}^\lambda\bigg\vert\chnlshort{p'}{}\right)\nonumber\\
&=&\sum_{\lambda=0}^{\infty}\sum_{m=-\lambda}^{\lambda}\frac{4\pi}{2\lambda+1}\frac{T_{pp'}^{\lambda m}}{r^{\lambda+1}}\langle\rsh{p}\vert X_{\lambda ,m}\vert\rsh{p'}\rangle,\label{ap1:eq:term1}
\end{eqnarray}
where $\langle\rsh{p}\vert X_{\lambda ,m}\vert\rsh{p'}\rangle$ is an integral over three real spherical harmonics that can be shown to always be proportional to a single standard Gaunt coefficient or zero~\cite{homeier1996}. Additionally, the integral over real spherical harmonics is invariant with respect to any permutation of the three pairs of the $\{l,m\}$ quantum numbers. The elements of the matrix $\mathbf{T}^{\lambda m}$ are given by the formula:
\begin{eqnarray}
T_{pp'}^{\lambda m}=\sum_{i=1}^{N}\langle\Phi_{p_{t}}\vert X_{\lambda ,m}(\hat{\mathbf{r}}_{i})r_{i}^\lambda\vert\Phi_{p_{t}'}\rangle.
\end{eqnarray}
The nuclear contribution (\ref{eq:We:nuc}) is evaluated similarly:
\begin{eqnarray}
&&W_{nuc,pp'}^{E}(r)=-\sum_{\lambda=0}^{\infty}\frac{1}{r^{\lambda+1}}\sum_{k=1}^{Nuclei}\bigg(\chnlshort{p}{}\bigg\vert Z_{k}P_{\lambda}(\hat{\mathbf{r}}_{N+1}.\hat{\mathbf{R}}_{k})R_{k}^{\lambda}\bigg\vert\chnlshort{p'}{}\bigg)=\nonumber\\
&=&-\langle \Phi_{p_{t}}\vert\Phi_{p_{t}'}\rangle \sum_{\lambda=0}^{\infty}\sum_{m=-\lambda}^{\lambda}\frac{4\pi}{2\lambda+1}\frac{1}{r^{\lambda+1}}\langle\rsh{p}\vert X_{\lambda ,m}\vert\rsh{p'}\rangle\nonumber\\
&\times&\sum_{k=1}^{Nuclei}Z_{k}X_{\lambda,m}(\hat{\mathbf{R}}_{k})R_{k}^{\lambda}.\label{ap1:eq:term2}
\end{eqnarray}
Collecting the electronic and the nuclear contributions (Eqns.~(\ref{ap1:eq:term1}) and (\ref{ap1:eq:term2})), we get the final expression for the coupling potentials $W_{pp'}^{E}(r)$:
\begin{eqnarray}
&&W_{pp'}^{E}(r)=\sum_{\lambda=0}^{\infty}\frac{1}{r^{\lambda+1}}\sqrt{\frac{4\pi}{2\lambda+1}}\sum_{m=-\lambda}^{\lambda}\langle\rsh{p}\vert X_{\lambda ,m}\vert\rsh{p'}\rangle\times\nonumber\\ 
&\times&\underbrace{\sqrt{\frac{4\pi}{2\lambda+1}}\left( T_{pp'}^{\lambda m} - \langle \Phi_{p_{t}}\vert\Phi_{p_{t}'}\rangle \sum_{k=1}^{Nuclei}Z_{k}X_{\lambda,m}(\hat{\mathbf{R}}_{k})R_{k}^{\lambda}\right)}_{Q_{pp'}^{\lambda m}},\label{ap1:eq:final}
\end{eqnarray}
which coincides with (\ref{ap1:eq:couplings}). The coefficients $Q_{pp'}^{\lambda m}$ are the spherical multipole permanent or transition moments corresponding to the channel electronic states $p_{t}$ and $p_{t}'$ of the residual molecule. In practice, the sum over $\lambda$ is truncated at a small value (typically $\lambda_{max} = 2$). The largest possible contributing multipole moment is given by the rules for coupling of three spherical harmonics: $\lambda_{max}=2l_{max}$, where $l_{max}$ is the largest continuum angular momentum included in the calculation.

The $\lambda=0$ moment $Q_{pp'}^{00} = (N-\sum_{k=1}^{Nuclei}Z_{k})\delta_{p,p'}$ accounts for the long-range Coulomb interaction between the residual molecule and the unbound electron. However, this trivial moment is not calculated by the UKRmol+ codes. Consequently, in the actual implementation the corresponding Coulomb potential $(N-\sum_{k=1}^{Nuclei}Z_{k})/r$ is included explicitly in the definition of $h_{p}(r)$ in Eq.~(\ref{eq:hp}), and the definitions~(\ref{eq:We:el}~-~\ref{eq:We:nuc}) of $W_{el}^{E}$ and $W_{nuc}^{E}$ are formally amended as follows
\begin{eqnarray}
\overline{W}_{el,pp'}^{E}(r) &=& W_{el,pp'}^{E}(r) - \frac{N}{r}\delta_{p,p'},\\
\overline{W}_{nuc,pp'}^{E}(r) &=& W_{nuc,pp'}^{E}(r) + \sum_{k=1}^{Nuclei}\frac{Z_{k}}{r}\delta_{p,p'},
\end{eqnarray}
which ensures that the contribution of $\lambda=0$ to the expression~(\ref{ap1:eq:couplings}) is zero, while the higher-order coefficients are intact, cf.~\cite{rmt_arb,burke}.

\subsection{$W^{D}$ potentials for laser-target interaction}
The potentials describing the interaction of the laser with the residual target are defined as the matrix elements:
\begin{eqnarray}
W^{D}_{pp'}(t) &=& \left(\chnlshort{p}{p}\vert D_{N}\vert\chnlshort{p'}{p'}\right) = \left(\chnlshort{p}{p}\vert \sum_{i=1}^{N} \mathbf{E}(t).\mathbf{r}_{i}\vert\chnlshort{p'}{p'}\right).\label{wddef}
\end{eqnarray}
Separating the integrations over the target's electrons and the unbound one, we obtain:
\begin{eqnarray}
W^{D}_{pp'}(t) &=& \mathbf{E}(t)\cdot\langle\tgt{p_{t}}\vert\sum_{i=1}^{N}\mathbf{r}_{i}\vert\tgt{p_{t}'}\langle\rsh{p}\vert\rsh{p'}\rangle=\nonumber\\
&=&\mathbf{E}(t)\cdot\mathbf{d}_{p_{t},p_{t}'}\delta_{l_{p},l_{p'}}\delta_{m_{p},m_{p'}},\label{wdeq}
\end{eqnarray}
where the term $\mathbf{E}(t)\cdot\mathbf{d}_{p_{t},p_{t}'}$ is the dot product between the $x$,$y$,$z$ components of the electric field and the corresponding components of the permanent or transition dipole moments for the $N$-electron target states in the channels $p$ and $p'$. We can see that this potential is diagonal in $\{l,m\}$, but if at least two target states of different symmetries are included it will couple wavefunctions of different total symmetries. 

%If the electric field is linearly polarized we can simplify equation~(\ref{wdeq}) as:
%\begin{eqnarray}
%W^{D}_{pp'}(t) &=& E(t)w^{d}_{pp'},\\
%w^{d}_{pp'}&=& \left(\hat{\mathbf{e}}.\mathbf{d}_{i,j}\right)\delta_{l_{p},l_{p'}}\delta_{m_{p},m_{p'}},
%\end{eqnarray}
%where $\hat{\mathbf{e}} = (e_{x},e_{y},e_{z})$ is the unit vector in the direction of the field. Therefore the $w^{d}_{pp'}$ terms containing the dot product $\hat{\mathbf{e}}.\mathbf{d}_{i,j}$ are computed only once (at the beginning of the calculation) in order to accommodate an arbitrary direction of the linearly polarized field in the $W^{D}$ potential.

\subsection{$W^{P}$ potentials for laser-electron interaction}

The last potential is the one describing the interaction of the unbound electron with the laser field:

\begin{eqnarray}
W^{P}_{pp'}(r,t) = \left(\chnlshort{p}{p}\vert D_{N+1}\vert\chnlshort{p'}{p'}\right) = \left(\chnlshort{p}{p}\vert \mathbf{E}(t).\mathbf{r}_{N+1}\vert\chnlshort{p'}{p'}\right).\label{wpdef}
\end{eqnarray}
Following the same steps as in the evaluation of the $W^{D}$ potential, we obtain:
\begin{eqnarray}
&W^{P}_{pp'}&(r,t) = \langle\tgt{p_{t}}\vert\tgt{p_{t}'}\rangle\mathbf{E}(t).\langle\rsh{p}\vert\mathbf{r}_{N+1}\vert \rsh{p'}\rangle=\nonumber\\
&=& \langle\tgt{p_{t}}\vert\tgt{p_{t}'}\rangle r\sqrt{\frac{4\pi}{3}}\nonumber\\
&\times\biggl[&E_{x}(t).\langle\rsh{p}\vert X_{1,1}\vert\rsh{p'}\rangle \nonumber\\ 
&+&E_{y}(t).\langle\rsh{p}\vert X_{1,-1} \vert\rsh{p'}\rangle\nonumber\\
&+&E_{z}(t).\langle\rsh{p}\vert {X}_{1,0}\vert\rsh{p'}\rangle\biggr],\label{wp}
\end{eqnarray}
where we took advantage of orthonormality of the target states and the form of the real spherical harmonics for $l=1$:
\begin{eqnarray}
X_{1,m}(\mathbf{r}) = 
\begin{cases} \sqrt{\frac{3}{4\pi}}\frac{y}{r} &,m = -1, \\ 
\sqrt{\frac{3}{4\pi}}\frac{z}{r} &,m = 0, \\ 
\sqrt{\frac{3}{4\pi}}\frac{x}{r} &,m = 1.
\end{cases}
\end{eqnarray}
The $W^{P}$ potential is diagonal in the target space and can couple wavefunctions of different total symmetries. 
%If the field is linearly polarized then the potential can be rewritten as:
%\begin{eqnarray}
%W^{P}_{pp'}(\mathbf{r},t) &=& E(t)r_{N+1}w^{p}_{pp'},\\
%w^{p}_{pp'}=\delta_{i,j}r_{N+1}&\sqrt{\frac{4\pi}{3}}&[e_{x}.\langle\rsh{p}\vert X_{1,1}\vert\rsh{p'}\rangle \nonumber\\ &+&e_{y}.\langle\rsh{p}\vert X_{1,-1} \vert\rsh{p'}\rangle\nonumber\\ &+&e_{z}.\langle\rsh{p}\vert {X}_{1,0}\vert\rsh{p'}\rangle],
%\end{eqnarray}
%where $\hat{\mathbf{e}} = (e_{x},e_{y},e_{z})$ is the unit vector in the direction of the field. The $w^{p}_{pp'}$ terms can be calculated only once in order to accommodate an arbitrary direction of the linearly polarized field in the $W^{P}$ potential.

We conclude that to construct the potential $\mathbf{V}(r,t)$ from Eq.~(\ref{eq:Vpot}) for an arbitrarily polarized field, we require the potential coupling coefficients $a_{pp'\lambda}$ from the field-free molecular calculation, the dipole transition moments for the residual molecular states and the coupling coefficients for three real spherical harmonics. All these coefficients are evaluated using UKRmol+ and stored in the \mbox{\texttt{molecular\_data}} file.

\section{Conventions in the atomic and molecular codes}\label{conventions}

The RMT suite uses data generated from atomic and molecular codes which have been developed separately over several decades. As a result, the codes use slightly different conventions for some quantities, and care was taken to implement them correctly:
\begin{itemize}
    \item In the atomic case, the one-electron wavefunctions are expanded in a basis of complex spherical harmonics defined in the Fano-Racah phase convention~\cite{burke}. In the molecular one, however, real spherical harmonics are used~\cite{homeier1996}. In RMT, this choice is reflected in the different angular basis (channels) for the unbound electron (see Section~\ref{execution}), and in the additional phase factor $\imath$ multiplying the atomic reduced dipole matrix elements and the long-range potential matrix elements. The Fano-Racah phase is implemented through the function \mbox{\texttt{field\_sph\_component}} from the module \mbox{\texttt{electric\_field}}.
    \item The definition of the boundary amplitudes in the expression for the R-matrix differs by a factor of $\sqrt{2}$ between the atomic (A) and the molecular (M) codes:
    \begin{eqnarray}
    R_{ij}(E) &=& \frac{1}{2a}\sum_{k}\frac{w_{ik}^{A}w_{jk}^{A}}{E_{k}-E},\\
    R_{ij}(E) &=& \frac{1}{a}\sum_{k}\frac{w_{ik}^{M}w_{jk}^{M}}{E_{k}-E}.
    \end{eqnarray}
    The RMT suite uses the atomic convention. Switching from the molecular to the atomic one is achieved by multiplying the molecular boundary amplitudes $w_{ik}^{M}$ by $\sqrt{2}$ when generating the input file \mbox{\texttt{molecular\_data}} using UKRmol+.
\end{itemize}


% ZM todo:
% DONE: refer to Burke's book or Greg and Daniel's paper for a description of the outer region potentials
% DONE: include the molecular outer region potentials in a separate appendix
% Include a description of the atomic/molecular conventions: spherical harmonics, Fano-Racah convention, boundary amplitudes (sqrt(2)) factor, etc.

\section*{\label{sec:references}References}
\bibliographystyle{elsarticle-num}
\bibliography{mybib}






% \end{thebibliography}

\end{document}